\documentclass[journal]{IEEEtran}
\usepackage[noadjust]{cite}
\usepackage[caption=false,font=footnotesize]{subfig}
\usepackage{stfloats}
\usepackage{url}
\usepackage[colorlinks=true,linkcolor=black,citecolor=blue,urlcolor=blue]{hyperref}
\usepackage{amssymb,amsmath,amsthm}
\usepackage{array}
\usepackage{amsfonts}
\usepackage{graphicx}
  \graphicspath{{./pics/}}
  \DeclareGraphicsExtensions{.pdf,.png}
\usepackage{xspace}
\usepackage{pgf}
\usepackage{tikz}
  \usetikzlibrary{arrows,shapes,fit,positioning}
  \usetikzlibrary{automata,backgrounds}
  \usetikzlibrary{decorations.pathmorphing}
  \usetikzlibrary{calc}
  \usetikzlibrary{positioning}
  \usetikzlibrary{decorations.markings}
  \usetikzlibrary{intersections}
\usepackage[T1]{fontenc}
\usepackage[latin1]{inputenc}
\usepackage[english]{babel}
\usepackage{booktabs}
  \newcommand{\ra}[1]{\renewcommand{\arraystretch}{#1}}
\usepackage{graphbox}
\usepackage{outlines}
\usepackage{algpseudocode}
\usepackage[linesnumbered,ruled,noend,noline]{algorithm2e}
  
\usepackage{fixmath}
\usepackage[export]{adjustbox}
\usepackage[textwidth=1.9cm,color=green!10,textsize=footnotesize]{todonotes}

\allowdisplaybreaks

\newtheorem{thm}{Theorem}
\newtheorem{cor}[thm]{Corollary}
\newtheorem{lem}[thm]{Lemma}
\newtheorem{remark}[thm]{Remark}
\newtheorem{defn}[thm]{Definition}
\newtheorem{exmp}[thm]{Example}
\newtheorem{prob}[thm]{Problem}

\DeclareMathOperator{\supN}{\rm supN}

\DeclareMathOperator{\pp}{|\!|\!|}
\newcommand{\eps}{\varepsilon}

\newcommand{\complclass}[1]{{\scshape #1}\xspace}
\newcommand{\PSpace}{\complclass{PSpace}}

\begin{document}

\title{Hierarchical Supervisory Control under Partial Observation: Normality\thanks{This work extends the material presented in part at WODES~2020~\cite{wodes2020}.}}%

\author{Jan Komenda and Tom{\' a}{\v s}~Masopust
  \thanks{J. Komenda and T. Masopust are with the Institute of Mathematics of the Czech Academy of Sciences, Prague, Czechia, and with the Faculty of Science, Palacky University Olomouc, Czechia. 
  Emails: {\tt komenda@ipm.cz}, {\tt tomas.masopust@upol.cz}.}%
\thanks{Supported by the M\v{S}MT INTER-EXCELLENCE project LTAUSA19098, by the GA\v{C}R grant GC19\mbox{-}06175J, and by RVO~67985840.}
}

\markboth{}{}

\maketitle

\begin{abstract}
  Conditions preserving observability of specifications between the plant and its abstraction are essential for hierarchical supervisory control of discrete-event systems under partial observation. Observation consistency and local observation consistency were identified as such conditions.
  To preserve normality, only observation consistency is required. Although observation consistency preserves normality between the levels for normal specifications, for specifications that are not normal, observation consistency is insufficient to guarantee that the supremal normal sublanguage computed on the low level and on the high level coincide. We define modified observation consistency, under which the supremal normal sublanguages of different levels coincide. 
  We show that the verification of (modified) observation consistency is \PSpace-hard for finite automata and undecidable for slightly more expressive models than finite automata. Decidability of (modified) observation consistency is an open problem. Hence we further discuss two stronger conditions that are easy to verify.
  Finally, we illustrate the conditions on an example of a railroad controller and on a case study of a part of an MRI scanner.
\end{abstract}
 

\section{Introduction}
  Organizing systems into hierarchical structures is a common engineering practice used to overcome the combinatorial state-space explosion. This technique has many applications across control theory and computer science, including manufacturing, robotics, and artificial intelligence~\cite{BaierM15,HubbardC02,TorricoC2002,CunhaC07,LiMR2004,NgoS18,SyllaLRD18,Raisch2005,GirardP2006,David-HenrietRH12}.

  Hierarchical supervisory control of discrete-event systems (DES) viewed as a two-level system decomposition was introduced by Zhong and Wonham~\cite{ZhongW1990}. The low-level plant (the system to be controlled) is restricted by a specification over a high-level alphabet. Based on the high-level abstraction of the low-level plant, the aim of hierarchical supervisory control is to synthesize a nonblocking and maximally permissive high-level supervisor suitable for a low-level implementation. This requirement is known as {\em hierarchical consistency}.

  To achieve hierarchical consistency, Zhong and Wonham provided a sufficient condition of {\em output control consistency}, and extended the framework to hierarchical coordination control~\cite{ZhongW1990b}. 
  Wong and Wonham~\cite{WW96} later developed an abstract hierarchical supervisory control theory that achieves hierarchical consistency by {\em control consistency\/} and {\em observer property}, and applied the theory to the Brandin-Wonham timed discrete-event systems~\cite{WongW96a}; see Chao and Xi~\cite{ChaoX03} for more details on control consistency conditions. 
 
  Schmidt et al.~\cite{KS} extended hierarchical supervisory control to decentralized systems, and Schmidt and Breindl~\cite{SB11} weakened the sufficient condition to achieve maximal permissiveness of high-level supervisors in hierarchical supervisory control under complete observation. 

  Fekri and Hashtrudi-Zad~\cite{fekri2009} considered hierarchical supervisory control under partial observation, but they used  the model of Moore automata and defined the concepts of controllable and observable events based on vocalization. Hence, they needed a specific definition of the low-level supervisor. 

  We adapt the Ramadge-Wonham framework for DES where a system $G$ is modeled as a (deterministic) finite automaton over a low-level alphabet $\Sigma$, and the abstraction is modeled as a projection from $\Sigma$ to a high-level alphabet $\Sigma_{hi} \subseteq \Sigma$. The behavior of the high-level (abstracted) plant $G_{hi}$ is the projection of the behavior of the low-level plant $G$ to the high-level alphabet $\Sigma_{hi}$. Given a high-level specification $K$ over $\Sigma_{hi}$ describing the required behavior of $G$ in terms of high-level events, the problem is to synthesize a nonblocking and maximally permissive supervisor $S_{hi}$ on the high level that, running in parallel with the low-level plant, behaves as a nonblocking and maximally permissive low-level supervisor.
  
  The above-mentioned concepts of {\em observer property}~\cite{WW96} and {\em output} or {\em local control consistency}~\cite{ZhongW1990,SB11} were developed to achieve hierarchical consistency for DES under complete observation. Therefore, they are insufficient to achieve hierarchical consistency under partial observation. 
  We addressed hierarchical supervisory control under partial observation and achieved hierarchical consistency under the condition that all observable events are high-level events~\cite{KM10}. Later, Boutin et al.~\cite{cdc-ecc2011} provided weaker and less restrictive conditions of {\em local observation consistency\/} and {\em observation consistency}.

  While observation consistency suffices to preserve normality between the levels~\cite{cdc-ecc2011}, it does not suffice to preserve the supremality of normal sublanguages of specifications that are neither observable nor normal (Example~\ref{counterexample1}).
  To overcome this issue, we define a concept of {\em modified observation consistency\/} (Definition~\ref{defMOC}), which preserves the supremality of normal sublanguages of specifications between the levels and ensures thus that the nonblocking and maximally permissive high-level supervisor, running in parallel with the low-level plant, behaves as the nonblocking and maximally permissive low-level supervisor (Theorem~\ref{supN} and Corollary~\ref{cor_supN}).
  Comparing observation consistency with modified observation consistency, the latter is stronger (Lemma~\ref{MOCstrongerOC}). 
  
  From the computational side, verifying (modified) observation consistency is a \PSpace-hard problem (Theorem~\ref{thm5}), while decidability is a challenging open problem.
  Although we do not solve decidability for finite automata, we solve the open problem negatively for a slightly more expressive model of one-turn deterministic pushdown automata (one-turn DPDAs, Theorem~\ref{undec1}). One-turn DPDAs are less expressive than pushdown automata discussed in the supervisory control literature in the context of controllability and synthesis as a generalization to systems with automatic synthesis~\cite{gri07,gri10,automatica2011b,Sreenivas93,SchmuckSRN16}.
 
  To tackle the decidability question, we provide an alternative definition of (modified) observation consistency that may be useful to show that a given plant does not satisfy (modified) observation consistency (Theorem~\ref{inclusionTHM} and Theorem~\ref{inclusionTHMmoc}).
  
  However, even if the verification of (modified) observation consistency turned out to be decidable for finite automata, the \PSpace-hardness result means that the verification would be a computational obstacle. From the practical point of view, it is therefore meaningful to consider abstractions that satisfy stronger and computationally easily-verifiable conditions, such as
  (i) all observable events are high-level events, or 
  (ii) all high-level events are observable.
  These conditions can be easily verified and, in a sense, ensured by a suitable choice of abstraction.
  In Section~\ref{sec:Practical Conditions}, we show that both conditions are stronger than (modified) observation consistency.
  
  We provide an illustrative example in Subsection~\ref{railEx01}, motivated by the railroad example of Alur~\cite{Alur}, and discuss a case study of a part of an MRI scanner~\cite{theunissen} in Section~\ref{section8}.
  
  In part, the work was presented at WODES 2020. Compared with the conference version~\cite{wodes2020} providing no proofs or only sketches, we provide full proof details and revise and correct some claims. In particular, we correct the statement of~\cite[Theorem~5]{wodes2020}, claiming that the verification of (modified) observation consistency is \PSpace-complete for NFAs. Unfortunately, our reasoning was incomplete, and hence decidability remains open. Theorem~\ref{thm5} states the correct claim and strengthens the result from NFAs to DFAs. We further revised the statement of~\cite[Theorem~11]{wodes2020}, stated here as Theorem~\ref{supN}. The results not presented in the conference version include the undecidability of the verification of (modified) observation consistency for one-turn DPDAs, and the illustrations of applicability of the results in the railroad example and in the case study.
  
  This paper is the first part focusing on normal supervisors. The upcoming papers will focus on (i) relatively-observable supervisors, and (ii) on the applications of hierarchical supervisory control in modular and coordination supervisory control~\cite{report22}.

\section{Preliminaries and Definitions}\label{section2}
  We assume that the reader is familiar with the basic notions and concepts of supervisory control~\cite{CL08}. For a set $A$, $|A|$ denotes the cardinality of $A$. For an alphabet (finite nonempty set) $\Sigma$, $\Sigma^*$ denotes the set of all finite strings over $\Sigma$; the empty string is denoted by $\eps$. A language $L$ is a subset of $\Sigma^*$. The prefix closure of $L$ is the set $\overline{L}=\{w\in \Sigma^* \mid \text{there is } v\in\Sigma^* \text{ such that } wv\in L\}$, and $L$ is prefix-closed if $L=\overline{L}$.

  A {\em projection\/} $R\colon \Sigma^* \to \Gamma^*$, where $\Gamma \subseteq \Sigma$ are alphabets, is a morphism for concatenation that is defined by $R(a) = \eps$ for $a\in \Sigma\setminus \Gamma$, and $R(a) = a$ for $a\in \Gamma$. The action of $R$ on a string $a_1a_2\cdots a_n$ is to remove all events that are not in $\Gamma$, i.\,e., $R(a_1a_2\cdots a_n)=R(a_1)R(a_2)\cdots R(a_n)$. The inverse image of a $w\in\Gamma^*$ under $R$ is the set $R^{-1}(w)=\{s\in \Sigma^* \mid R(s) = w\}$. The definitions can readily be extended to languages. 

  A {\em nondeterministic finite automaton\/} (NFA) is a quintuple $G = (Q,\Sigma,\delta,q_0,F)$, where $Q$ is a finite set of states, $\Sigma$ is an alphabet, $q_0 \in Q$ is the initial state, $F\subseteq Q$ is the set of marked states, and $\delta \colon Q\times\Sigma \to 2^Q$ is the transition function that can be extended to the domain $2^Q\times \Sigma^*$ in the usual way. 
  Automaton $G$ is {\em deterministic\/} (DFA) if $|\delta(q,a)|\le 1$ for every state $q \in Q$ and every event $a \in \Sigma$, in which case we view the transition function $\delta$ as $\delta\colon Q\times \Sigma \to Q$.
  The language {\em generated\/} by $G$ is the set $L(G) = \{w\in \Sigma^* \mid \delta(q_0,w)\in Q \}$, and the language {\em marked\/} by $G$ is the set $L_m(G) = \{w\in \Sigma^* \mid \delta(q_0,w) \in F\}$.
  By definition, $L_m(G)\subseteq L(G)$, and $L(G)$ is prefix-closed. If $\overline{L_m(G)} = L(G)$, then $G$ is called {\em nonblocking}.

  A discrete-event system (DES) over $\Sigma$ is a DFA over $\Sigma$ together with the specification of {\em controllable events} $\Sigma_c$ and {\em uncontrollable events} $\Sigma_{uc} = \Sigma \setminus \Sigma_c$, and {\em observable events} $\Sigma_o$ and {\em unobservable events} $\Sigma_{uo} = \Sigma \setminus \Sigma_o$.
  
  The parallel composition of languages $L_i\subseteq \Sigma_i^*$ is the language $\|_{i=1}^{n} L_i = \cap_{i=1}^{n} P_i^{-1}(L_i)$, where $P_i \colon (\cup_{i=1}^{n} \Sigma_i)^* \to \Sigma_i^*$ is the projection, for $i=1,\ldots,n$. A corresponding definition of the parallel composition for automata can be found in the literature~\cite{CL08}. In particular, for NFAs $G_i$, we have $L(\|_{i=1}^{n} G_i) = \|_{i=1}^{n} L(G_i)$ and $L_m(\|_{i=1}^{n} G_i) = \|_{i=1}^{n} L_m(G_i)$. The languages $L_i$ are {\em (synchronously) nonconflicting\/} if $\overline{\|_{i=1}^{n} L_{i}} = \|_{i=1}^{n} \overline{L_{i}}$.

\section{Hierarchical Supervisory Control}\label{section3}
\begin{figure}
    \centering
    \includegraphics[scale=.8]{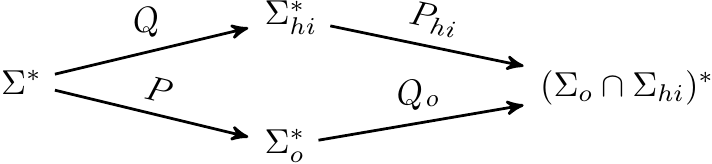}
    \caption{Commutative diagram of abstractions and projections.}
    \label{projections}
  \end{figure}

  Before stating the hierarchical supervisory control problem for partially observed DES, we review and fix the notation of the supervisory control theory~\cite{CL08}.
  We denote the low-level alphabet by $\Sigma$, the high-level alphabet by $\Sigma_{hi}\subseteq \Sigma$, the set of observable events by $\Sigma_o\subseteq \Sigma$, the system's partial observation by projection $P\colon \Sigma^* \to \Sigma^*_o$, the high-level abstraction by projection $Q\colon \Sigma^* \to \Sigma_{hi}^*$, and the corresponding restricted observations and abstractions by
  $P_{hi}\colon \Sigma_{hi}^{*} \to (\Sigma_{hi}\cap \Sigma_o)^*$ and
  $Q_o\colon \Sigma_o^* \to (\Sigma_{hi}\cap \Sigma_o)^*$, cf. Figure~\ref{projections}.

  For a DES $G$ over $\Sigma$, we denote by $\Gamma = \{\gamma \subseteq \Sigma \mid \Sigma_{uc} \subseteq \gamma\}$ the set of control patterns. A {\em supervisor\/} of $G$ with respect to $\Gamma$ is a map $S\colon P(L(G)) \to \Gamma$. The {\em closed-loop system\/} of $G$ and $S$ is the minimal language $L(S/G)$ such that $\eps \in L(S/G)$ and for every $s\in L(S/G)$, if $sa \in L(G)$ and $a \in S(P(s))$, then $sa \in L(S/G)$. Intuitively, the supervisor disables some of the transitions, but never an uncontrollable transition. 

  Considering the marked language of the closed-loop system, there are two approaches: (i) the marking is adopted from the plant $G$, that is, $L_m(S/G) = L(S/G) \cap L_m(G)$, and (ii) the supervisor marks according to a given specification $M\subseteq L(G)$, i.\,e., $L_m(S/G)=L(S/G)\cap M$. 
  In the latter case, the existence of a supervisor that achieves the specification is equivalent to controllability and observability of the specification, whereas, in the former case, an additional assumption of $L_m(G)$\mbox{-}closedness is needed~\cite[Section~6.3]{Won04}. 
  If the closed-loop system is nonblocking, i.\,e., $\overline{L_m(S/G)}=L(S/G)$, then the supervisor $S$ is called {\em nonblocking}.
   
  In this paper, we focus on the preservation of controllability and observability between the levels, which is equivalent to considering marking supervisors.
  
  \begin{prob}[Hierarchical Supervisory Control]\label{prob1}
    Let $G$ be a low-level plant over $\Sigma$, and let $K$ be a high-level specification over $\Sigma_{hi} \subseteq \Sigma$. We define the abstracted high-level plant $G_{hi}$ over $\Sigma_{hi}$ by $L(G_{hi}) = Q(L(G))$ and $L_m(G_{hi}) = Q(L_m(G))$. Based on $G_{hi}$ and $K$, the hierarchical supervisory control problem is to find theoretical conditions, under which a nonblocking low-level supervisor $S$ exists such that $L_m(S/G) = K \| L_m(G)$. 
    \hfill$\diamond$
  \end{prob}
 
  Intuitively, we look for conditions on the low-level plant $G$, under which controllability and observability of $K\| L_m(G)$ with respect to $L(G)$ is equivalent to controllability and observability of $K$ with respect to $L(G_{hi})$.
  Such conditions are known for controllability, namely {\em $L_m(G)$-observer}~\cite{WW96} and {\em output} or {\em local control consistency}~\cite{ZhongW1990,SB11}, but not yet well understood for observability. Although Boutin et al.~\cite{cdc-ecc2011} identified such conditions, observation consistency and local observation consistency, the decidability status of their verification is open for finite automata, and the conditions guarantee neither the preservation of maximal permissiveness of supervisors between the levels, nor the preservation of the supremality of normality between the levels, see Section~\ref{section5}.

  We now recall the definitions of observation consistency and local observation consistency of Boutin et al.~\cite{cdc-ecc2011}.
  
  A prefix-closed language $L \subseteq \Sigma^*$ is {\em observation consistent} (OC) with respect to projections $Q$, $P$, and $P_{hi}$ if for every high-level strings $t,t' \in Q(L)$ with $P_{hi}(t) = P_{hi}(t')$, there are low-level strings $s,s' \in L$ such that $Q(s) = t$, $Q(s') = t'$, and $P(s) = P(s')$. Intuitively, any two strings with the same observation in the high-level plant have corresponding strings with the same observation in the low-level plant.
  
  A prefix-closed language $L \subseteq \Sigma^*$ is {\em locally observation consistent} (LOC) with respect to $Q$, $P$, and $\Sigma_c$ if, for every $s,s'\in L$ and every $e\in \Sigma_c \cap \Sigma_{hi}$ such that $Q(s)e, Q(s')e \in Q(L)$ and $P(s) = P(s')$, there are low-level strings $u,u'\in (\Sigma\setminus \Sigma_{hi})^*$ such that $P(u) = P(u')$ and $sue, s'u'e\in L$. Intuitively, if we can extend two observationally equivalent high-level strings by the same controllable event, we can also extend their corresponding low-level observationally equivalent strings by this same event in the original plant after (possibly empty) low-level strings with the same observations. 
  
  In this paper, we are interested only in the OC condition, because, as shown in (3) of Theorem~\ref{boutinThm}, the LOC condition plays no role in preserving normality. We discuss the OC condition in detail in Section~\ref{appendixB}. 

  Before we summarize the main results of Boutin et al.~\cite{cdc-ecc2011} in Theorem~\ref{boutinThm}, we recall the basic concepts of supervisory control that are necessary for their understanding.

  Let $G$ be a DES over $\Sigma$. A language $K\subseteq L_m(G)$ is {\em controllable} with respect to $L(G)$ and uncontrollable events $\Sigma_{uc}$ if $\overline{K}\Sigma_{uc}\cap L(G)\subseteq \overline{K}$, and 
  $K$ is {\em normal\/} with respect to $L(G)$ and $P\colon\Sigma^*\to\Sigma_o^*$ if $\overline{K} = P^{-1}[P(\overline{K})]\cap L(G)$~\cite{LinWon88}.
  A language $K\subseteq L(G)$ is {\em observable\/} with respect to $L(G)$, observable events $\Sigma_o$ with the corresponding projection $P\colon \Sigma^*\to \Sigma_o^*$, and controllable events $\Sigma_c$ if for every $s,s'\in L(G)$ with $P(s)=P(s')$ and every $e\in \Sigma_c$, whenever $se \in \overline{K}$, $s'e \in L(G)$, and $s' \in \overline{K}$, we have $s'e \in \overline{K}$. 
  For algorithms verifying controllability and observability, see the literature~\cite{CL08}.
  
  Projection $Q\colon \Sigma^* \to \Sigma_{hi}^*$ is an {\em $L_m(G)$-observer} for a nonblocking DES $G$ over $\Sigma$ if, for every $t\in Q(L_m(G))$ and $s\in \overline{L_m(G)}$, whenever $Q(s)$ is a prefix of $t$, then there is $u\in \Sigma^*$ such that $su\in L_m(G)$ and $Q(su)=t$.

  We say that $Q$ is {\em locally control consistent\/} (LCC) for $s\in L(G)$ if, for every $e \in \Sigma_{hi} \cap \Sigma_{uc}$ such that $Q(s)e \in L(G_{hi})$, either there is no $u\in (\Sigma \setminus \Sigma_{hi})^*$ such that $sue \in L(G)$ or there is $u\in (\Sigma_{uc}\setminus \Sigma_{hi})^*$ such that $sue\in L(G)$. We call $Q$ LCC for a language $M\subseteq L(G)$ if $Q$ is LCC for every $s\in M$.

  \begin{thm}[Boutin et al.~\cite{cdc-ecc2011}]\label{boutinThm}
    Let $G$ be a nonblocking DES over $\Sigma$, and let $K\subseteq Q(L_m(G))$ be a high-level specification.

    (1) If $L(G)$ is observation consistent w.r.t. $Q$, $P$, and $P_{hi}$, $K$ and $L_m(G)$ are nonconflicting, and $L(G)$ is locally observation consistent w.r.t. $Q$, $P$, and $\Sigma_{c}$, then $K$ is observable w.r.t. $Q(L(G))$, $\Sigma_{hi} \cap \Sigma_o$, and $\Sigma_{hi} \cap \Sigma_c$ if and only if $K \| L_m(G)$ is observable w.r.t. $L(G)$, $\Sigma_o$, and $\Sigma_c$.
      
    (2) If $Q$ is an $L_m(G)$-observer and LCC for $L(G)$, and $L(G)$ is observation consistent w.r.t. $Q$, $P$, and $P_{hi}$, and locally observation consistent w.r.t. $Q$, $P$, and $\Sigma_c$, then $K$ is controllable w.r.t. $Q(L(G))$ and $\Sigma_{uc} \cap \Sigma_{hi}$, and observable w.r.t. $Q(L(G))$, $\Sigma_o \cap \Sigma_{hi}$, and $\Sigma_c \cap \Sigma_{hi}$ if and only if $K\|L_m(G)$ is controllable w.r.t. $L(G)$ and $\Sigma_{uc}$, and observable w.r.t. $L(G)$, $\Sigma_o$, and $\Sigma_c$.
      
    (3) If $L(G)$ is observation consistent w.r.t. $Q$, $P$, and $P_{hi}$, and $K$ and $L_m(G)$ are nonconflicting, then $K$ is normal w.r.t. $Q(L(G))$ and $P_{hi}$ if and only if $K\| L_m(G)$ is normal w.r.t. $L(G)$ and $P$.
    \hfill\IEEEQEDhere
  \end{thm}

\section{Observation Consistency}\label{section4}
\label{appendixB}
  In this section, we express the OC condition as an inclusion of relational languages. This expression may be useful to show that a given plant does not satisfy OC, as illustrated by Example~\ref{ex:illustration} below. To this end, we generalize the parallel composition to an arbitrary set of synchronizing events using the event pairs in a similar way as discussed, e.\,g., in Arnold~\cite{Arnold94}.
  
  \begin{defn}\label{soucin}
    Let $L_1\subseteq \Sigma_1^*$ and $L_2\subseteq \Sigma_2^*$ be the languages of the NFAs $G_1=(Q_1,\Sigma_1,\delta_1,q_1,F_1)$ and $G_2=(Q_2,\Sigma_2,\delta_2,q_2,F_2)$, respectively. Let $\Sigma' \subseteq \Sigma_1 \cap \Sigma_2$ be a set of synchronizing events. The parallel composition of $L_1$ and $L_2$ synchronized on the events of $\Sigma'$ is denoted by $L_1 \pp_{\Sigma'} L_2$ and defined as the language of the NFA
    \[
      G_1 \pp_{\Sigma'} G_2 = (Q_1\times Q_2, \Gamma, \delta, (q_1,q_2), F_1\times F_2)
    \]
    where the alphabet $\Gamma \subseteq (\Sigma_1\cup\{\eps\}) \times (\Sigma_2\cup\{\eps\})$ is a set of pairs of events based on the synchronization alphabet $\Sigma'$. There are two different categories of pairs to construct, corresponding to events in $\Sigma'$ and events in $(\Sigma_1\cup\Sigma_2)\setminus\Sigma'$. For $a\in \Sigma'$ we have the pair $(a,a)$ in $\Gamma$, for $a\in \Sigma_1\setminus \Sigma'$ we have the pair $(a,\eps)$ in $\Gamma$, and for $a\in \Sigma_2\setminus \Sigma'$ we have the pair $(\eps,a)$ in $\Gamma$. The transition function $\delta\colon (Q_1\times Q_2) \times \Gamma \to 2^{Q_1\times Q_2}$ is defined on any $(p,q)\in Q_1\times Q_2$ as follows:
    \begin{outline}
      \1 for $a\in\Sigma'$, $G_1$ and $G_2$ proceed synchronously, i.\,e.,
        \2 $\delta((p,q),(a,a)) = \delta_1(p,a) \times \delta_2(q,a)$;
      \1 for $a\in\Sigma_1\setminus\Sigma'$, $G_2$ does not move, i.\,e.,
        \2 $\delta((p,q),(a,\eps)) = \delta_1(p,a)\times \{q\}$; 
      \1 for $a\in\Sigma_2\setminus\Sigma'$, $G_1$ does not move, i.\,e.,
        \2 $\delta((p,q),(\eps,a)) = \{p\} \times \delta_2(q,a)$;
      \1 undefined otherwise. \hfill$\diamond$
    \end{outline}
  \end{defn}
    
  For simplicity, we write a sequence of event pairs, such as $(a_1,\eps)(a_2,a_2)(\eps,a_3)$, as a pair of concatenated components, that is, $(a_1a_2,a_2a_3)$. In accordance with Definition~\ref{soucin}, the language of $G_1 \pp_{\Sigma'} G_2$ consists of all pairs of strings $(w,w') \in L_1 \times L_2$, where $w$ and $w'$ coincide on the events of $\Sigma'$, that is, $P'(w)=P'(w')$ for the projection $P'\colon (\Sigma_1\cup\Sigma_2)^* \to \Sigma'^*$. We call such languages {\em relational languages}. 
  
  We now use this structure to express the OC condition as an inclusion of two relational languages.
  
  \begin{thm}\label{inclusionTHM}
    Let $L \subseteq \Sigma^*$ be a prefix-closed language, and let $\Sigma_o$ and $\Sigma_{hi}$ be resp.~the observable and high-level alphabets. Then, $L$ is OC with respect to $Q$, $P$, and $P_{hi}$ if and only if
    \begin{align}\label{OCinclusion}
      Q(L)\pp_{\Sigma_{hi}\cap \Sigma_o} Q(L) 
      \subseteq 
      Q\left(L\pp_{\Sigma_o} L\right)
    \end{align}
    where, for an event $(a,b)$, $Q(a,b)=(Q(a),Q(b))$. 
  \end{thm}
  \begin{IEEEproof}
    Recall that OC states that for all $t,t' \in Q(L)$ with $P_{hi}(t) = P_{hi}(t')$, there are $s,s' \in L$ such that $Q(s) = t$, $Q(s') = t'$, and $P(s) = P(s')$. The intuition behind \eqref{OCinclusion} is to couple all strings $t,t'\in Q(L)$ with the same high-level observations\mbox{---}the pairs $(t,t')\in Q(L)\pp_{\Sigma_{hi}\cap \Sigma_o} Q(L)$\mbox{---}and to verify that for every such pair there are strings $s,s' \in L$ with the same observations\mbox{---}the pairs $(s,s')\in L\pp_{\Sigma_o} L$\mbox{---}that are abstracted to the pair $(t,t')$, i.\,e., $(Q(s),Q(s'))=(t,t')$. 
    
    Formally, we first show that if $L$ is OC, then \eqref{OCinclusion} holds. To this end, assume that $(t,t')\in Q(L) \pp_{\Sigma_{hi}\cap\Sigma_o} Q(L)$. By the definition of $\pp_{\Sigma_{hi}\cap\Sigma_o}$, we have that $t,t' \in Q(L)$ and $t,t'$ coincide on the events of $\Sigma_{hi}\cap\Sigma_o$, i.\,e., $P_{hi}(t) = P_{hi}(t')$. Since $L$ is OC, there are $s,s' \in L$ such that $Q(s) = t$, $Q(s') = t'$, and $P(s) = P(s')$. However, $P(s)=P(s')$ implies that $(s,s')\in L\pp_{\Sigma_o} L$, and $Q(s)=t$ and $Q(s')=t'$ imply that $(Q(s),Q(s'))=(t,t')$, which shows inclusion \eqref{OCinclusion}.
    
    On the other hand, assume that \eqref{OCinclusion} holds. We show that $L$ is OC. To do this, consider $t,t'\in Q(L)$ such that $P_{hi}(t) = P_{hi}(t')$. By the definition of $\pp_{\Sigma_{hi}\cap\Sigma_o}$, we obtain that $(t,t')\in Q(L)\pp_{\Sigma_{hi}\cap\Sigma_o} Q(L)$. Since \eqref{OCinclusion} holds, $(t,t')\in Q(L \pp_{\Sigma_o} L)$, which means that there is a pair $(s,s')\in L\pp_{\Sigma_o} L$ such that $(Q(s),Q(s'))=(t,t')$. However, $(s,s')\in L\pp_{\Sigma_o} L$ implies that the strings $s$ and $s'$ belong to $L$ and coincide on the events from $\Sigma_o$, i.\,e., $P(s)=P(s')$. Therefore, $L$ is OC.
  \end{IEEEproof}

  In the following example, we illustrate the previous construction and its application for finding counterexamples violating the OC condition.
  
  \begin{exmp}\label{ex:illustration}
    Consider the plant over $\Sigma=\{a,b,c\}$ defined by the automaton $G$ depicted in Figure~\ref{figIllustrationA} (left), and denote its language by $L$. Let $\Sigma_{hi}=\{a,b\}$ be the high-level events, and let $\Sigma_o=\{b,c\}$ be the observable events. The abstracted plant with the language $Q(L)$ is depicted in Figure~\ref{figIllustrationA} (right). 
    \begin{figure}
      \centering
      \includegraphics[align=c,scale=.75]{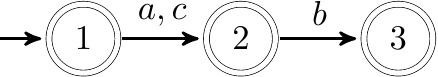}\qquad
      \includegraphics[vshift=-.22cm,scale=.75]{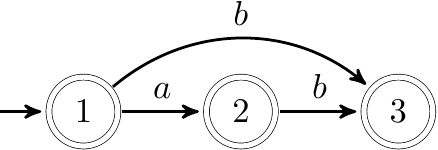}
      \caption{Plant $G$ with language $L$ and the plant of its abstraction $Q(L)$.}
      \label{figIllustrationA}
    \end{figure}
    The compositions $Q(L) \pp_{\{b\}} Q(L)$, $L\pp_{\{b,c\}} L$, $Q(L\pp_{\{b,c\}} L)$ are depicted in Figure~\ref{figIllustrationB}. Notice that the pair of strings $(ab,b)$ belongs to $Q(L) \pp_{\{b\}} Q(L)$ but not to $Q(L\pp_{\{b,c\}} L)$, and hence the strings $ab$ and $b$ violate the OC condition, showing thus that $L$ is not OC. 
    \hfill$\diamond$ 
  \end{exmp}

  \begin{figure}
    \centering
    \includegraphics[align=c,scale=.75]{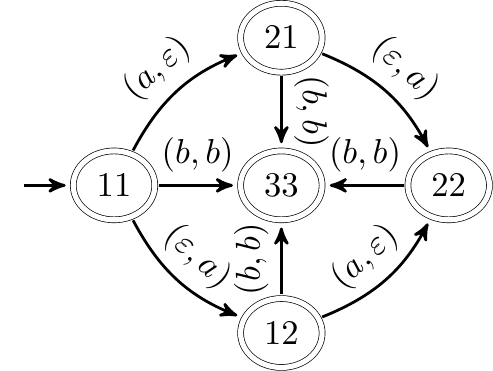}
    \includegraphics[align=c,scale=.75]{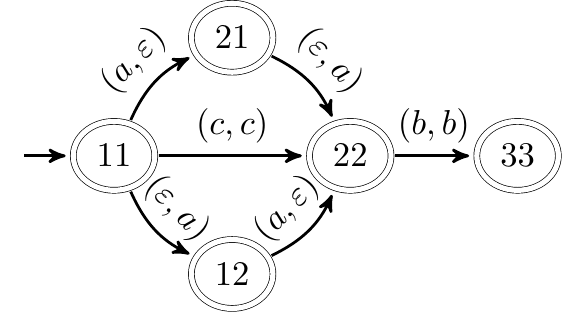}
    \includegraphics[align=c,scale=.75]{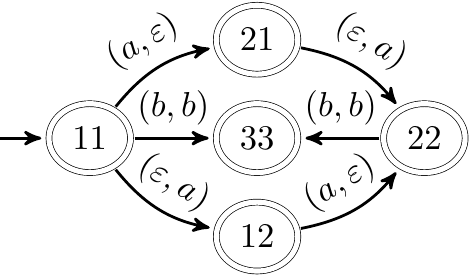}
    \caption{Languages $Q(L)\pp_{\{b\}} Q(L)$, $L\pp_{\{b,c\}} L$, and $Q(L\pp_{\{b,c\}} L)$, resp.}
    \label{figIllustrationB}
  \end{figure}

  Since both sides of inclusion~\eqref{OCinclusion} are represented by finite automata, it could seem that we can verify the OC condition by checking the inclusion by the classical algorithms verifying the inclusion of the languages of two NFAs. However, it is not the case because the languages are not classical languages, but rather relational languages. 

  The following example illustrates this issue.
  \begin{exmp}\label{ex:illustration2}
    Consider the plant defined by the DES $G$ depicted in Figure~\ref{figIllustration2A} (left), and denote its language by $L$.
    Let $\Sigma_{hi}=\{a,c\}$ be the high-level alphabet, and let $\Sigma_o=\{b\}$ be the single observable event. 
    \begin{figure}
      \centering
      \includegraphics[align=c,scale=.75]{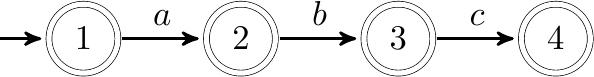}\quad
      \includegraphics[align=c,scale=.75]{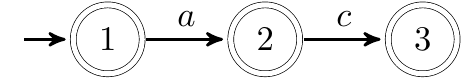}
      \caption{Plant $G$ with the language $L$ and its abstraction $Q(L)$.}
      \label{figIllustration2A}
    \end{figure}
    The languages $Q(L)\pp_{\emptyset} Q(L)$, $L\pp_{\{b\}} L$, and $Q(L\pp_{\{b\}} L)$ are shown in Figure~\ref{figIllustration2B}.
    The reader may see that the string $(a,\eps)(c,\eps)(\eps,a)(\eps,c)\in Q(L) \pp_{\emptyset} Q(L)$ but not in $Q(L\pp_{\{b\}} L)$, which seems to suggest that inclusion~\eqref{OCinclusion} does not hold. However, $(a,\eps)(c,\eps)(\eps,a)(\eps,c)=(ac,ac)=(a,\eps)(\eps,a)(c,\eps)(\eps,c)$, and the string $(a,\eps)(\eps,a)(c,\eps)(\eps,c) \in Q(L\pp_{\{b\}} L)$. Therefore, the pair of strings $(ac,ac)$ belongs to both sides of inclusion~\eqref{OCinclusion} and, in fact, the inclusion \mbox{holds. \hfill$\diamond$} 
  \end{exmp}

  \begin{figure}
    \centering
    \includegraphics[align=c,scale=.64]{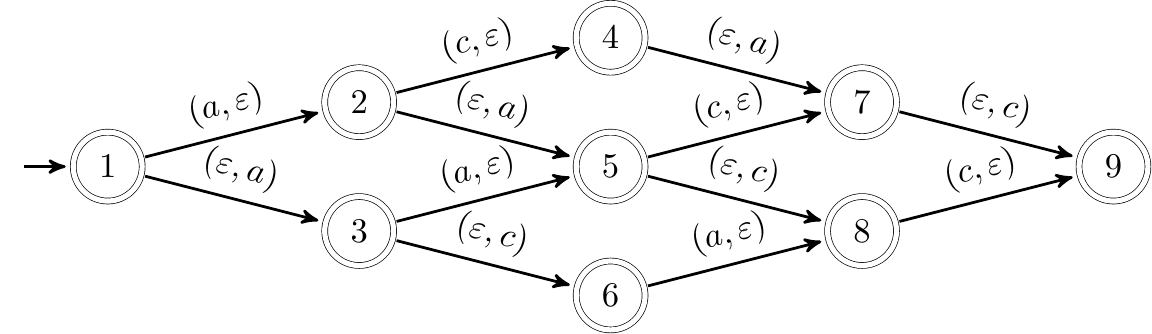}\\\medskip
    \includegraphics[align=c,scale=.64]{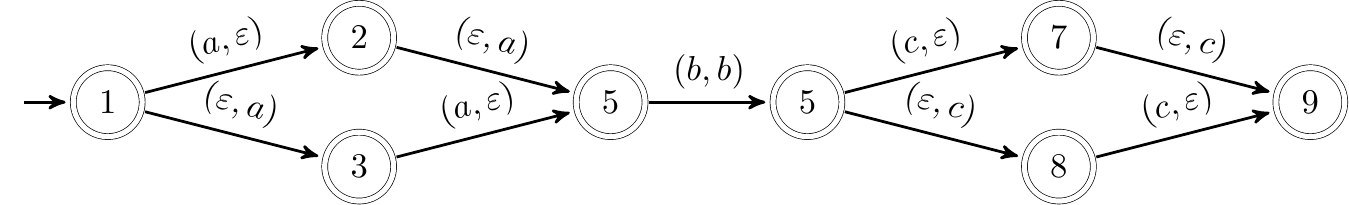}\\\medskip
    \includegraphics[align=c,scale=.64]{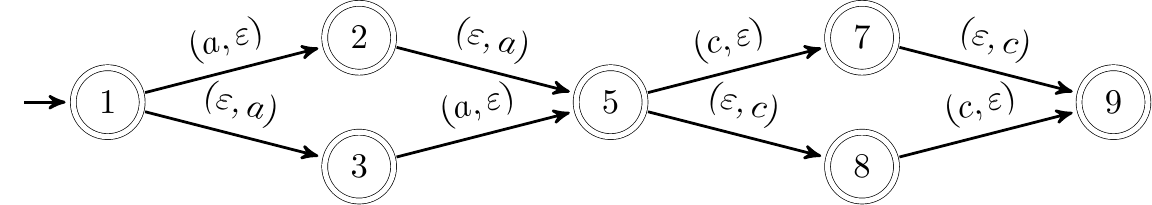}
    \caption{Languages $Q(L)\pp_{\emptyset} Q(L)$, $L\pp_{\{b\}} L$, and $Q(L\pp_{\{b\}} L)$, resp.}
    \label{figIllustration2B}
  \end{figure}
  
  Considering the previous example, the reader may notice that for every string from the language $Q(L)\pp_{\emptyset} Q(L)$ that is suspected of violating the OC condition, such as the string $(a,\eps)(c,\eps)(\eps,a)(\eps,c)$, we can check whether at least one of its permutations is present in the language $Q(L\pp_{\{b\}} L)$. If none of the permutations belongs to $Q(L\pp_{\{b\}} L)$, the OC condition does not hold. 
  Specifically, we may notice that the plant $G$ preserves the order of $a$ and $c$, and hence we only need to consider permutations that preserve the order of $a$ and $c$. This means that the pairs of events $((a,\eps),(\eps,c))$ and $((\eps,a),(c,\eps))$ may be interchanged in the strings. Adding all strings constructed this way to $Q(L\pp_{\{b\}} L)$ results in the commutative closure of $Q(L\pp_{\{b\}} L)$, which, in the previous example, coincides with $Q(L)\pp_{\emptyset} Q(L)$, and hence the inclusion $Q(L)\pp_{\emptyset} Q(L) \subseteq Q(L\pp_{\{b\}} L)$ holds.
  
  To generalize this observation, let $\Sigma$ be an alphabet, and let $I\subseteq \Sigma \times \Sigma$ be a symmetric relation. The closure of a language $L\subseteq \Sigma^*$ with respect to the partial commutation relation $I$ is denoted by $[L]_I$. The language $[L]_I$ is defined as the least language over $\Sigma$ that contains $L$ and satisfies that $uabv \in [L]_I$ if and only if $ubav \in [L]_I$ for every $u,v\in \Sigma^*$ and $(a,b)\in I$.
  It is known that regular languages are not closed under partial commutation, and that the question whether the closure of a regular language remains regular or not is decidable if and only if the partial commutative relation $I$ is transitive~\cite{Sakarovitch92}.\footnote{Here the notion of transitivity has the meaning that for every three {\em different\/} events $a,b,c\in \Sigma$, if $(a,b),(b,c)\in I$, then also $(a,c)\in I$.} 
  For the previous example, the partial commutative relation $I=\{((a,\eps),(\eps,c)),((\eps,c),(a,\eps)),((\eps,a),(c,\eps)),((c,\eps),(\eps,a))\}$ is transitive in the considered sense.
  More details can be found in the literature~\cite{GomezGP08,MuschollP96}. Consequently, in these cases, if the closure is regular and can be algorithmically constructed, the OC condition is decidable by the classical NFA inclusion algorithms. In general, however, the decidability status of the verification of OC is open, as well as the complexity of the decidable cases, see further discussion in Section~\ref{decMOC}.

\section{Supremality of Normal Sublanguages}\label{section5}\label{secPresSupN}
  The hierarchical supervisory control problem requires that the specification is exactly achieved by a supervisor. In other words, the specification is observable. What if the specification is not observable? A common approach is to find a suitable observable sublanguage of the specification and to construct a supervisor to achieve this sublanguage. 
  Since there are no supremal observable sublanguages in general, the supremal normal sublanguages are considered instead. 
 
  \begin{prob}[Hierarchical Supervisory Control Synthesis]
  \label{prob2}
    Let $G$ be a low-level plant over $\Sigma$, and let $K$ be a high-level specification over $\Sigma_{hi} \subseteq \Sigma$. Define the abstracted high-level plant $G_{hi}$ over $\Sigma_{hi}$ by $L(G_{hi}) = Q(L(G))$ and $L_m(G_{hi}) = Q(L_m(G))$. The hierarchical supervisory control synthesis problem is to construct, based on the high-level plant $G_{hi}$ and the high-level specification $K$, a nonblocking low-level supervisor $S$---one that is maximally permissive to the extent bounded by a mathematically well-behaved concept stronger than observability---such that $L_m(S/G) \subseteq K\| L_m(G)$.\footnote{In this paper, normality is the stronger observability concept applied.}
    \hfill$\diamond$
  \end{prob}

  Intuitively, we want to construct a nonblocking and maximally permissive high-level supervisor $S_{hi}$ with respect to the high-level plant $G_{hi}$ and the high-level specification $K$ satisfying $L_m(S_{hi}/G_{hi}) \subseteq K$ without constructing the nonblocking and maximally permissive low-level supervisor $S$, such that the closed-loop system $L_m(S_{hi}/G)$ coincides with the closed-loop system $L_m(S/G)\subseteq K\|L_m(G)$. Denoting the automaton realizing the high-level supervisor $S_{hi}$ by $G_{S_{hi}}$, we may implement the low-level supervisor in the form of $L_m(S/G) = L_m(G_{S_{hi}} \| G) = L_m(S_{hi}/G)$. In other words, we only construct the high-level supervisor, obtaining the closed-loop system $L_m(S/G)$ as the closed-loop system $L_m(S_{hi}/G)$.
  
  In the sequel, we focus only on the construction of normal supervisors, because the conditions for controllable supervisors are well known in the literature as discussed above. 
  
  Compared with (3) of Theorem~\ref{boutinThm}, which states that, under the OC condition, the high-level specification $K$ is normal if and only if the low-level language $K\| L_m(G)$ is normal, the following example shows that OC is insufficient to preserve the supremality of normality if the supremal normal sublanguage of $K$ is a strict subset of $K$.
  The problem is that the supremal normal sublanguage of $K\| L_m(G)$ is in general not of the form $X\| L_m(G)$ for a convenient language $X\subseteq K$ that would be the supremal normal sublanguage of $K$ at the high level.

  For a prefix-closed language $L$ and a specification $K\subseteq L$, we denote by $\supN(K,L,P)$ the supremal normal sublanguage of $K$ with respect to the plant language $L$ and the projection $P$ to observable events.

  \begin{exmp}\label{counterexample1}
    Consider the alphabet $\Sigma=\{a,b,c\}$ with observable events $\Sigma_o=\{a,c\}$ and high-level events $\Sigma_{hi}=\{b,c\}$, and the language $L=\{\eps,a,b,c,ba,ac,bac\}$. To show that $L$ is OC, notice that $P_{hi}(\eps)=\eps=P_{hi}(b)$ and $P_{hi}(c)=c=P_{hi}(bc)$. There are two cases: (i) $t=\eps$ and $t'=b$, which is trivial because we can choose $s=t=\eps$ and $s'=t'=b$ to satisfy OC, and (ii) $t=c$ and $t'=bc$, where we choose $s=ac$ and $s'=bac$, since then $Q(s)=c=t$, $Q(s')=bc=t'$, and $P(s)=ac=P(s')$. Thus, $L$ is OC.
    
    To compute the supremal normal sublanguage of $K=\{\eps,b,c\}\subseteq Q(L)=\{\eps,b,c,bc\}$, we use the formula 
    \[
      \supN(B,M,P)=B-P^{-1}P(M-B)\Sigma^*
    \]
    for prefix-closed languages $B\subseteq M \subseteq \Sigma^*$~\cite{brandt}, and obtain that
      $K\| L = a^*ba^* \cup a^*ca^* \cup a^* \cap L =\{\eps,a,b,c,ba,ac\}$,
      $L-K\| L=\{bac\}$, and
      $P^{-1}P(bac)=P^{-1}(ac)=b^*ab^*cb^*$,
    i.\,e.,
    $
      c\in \supN(K\| L,L,P) 
        = K\| L -P^{-1}P(L-K\|L)\Sigma^* = \{\eps,a,b,c,ba\} 
    $.
    On the other hand,
      $Q(L)-K=\{\eps,b,c,bc\}-\{\eps,b,c\}=\{bc\}$,
      $P_{hi}(bc)=c$, and 
      $P_{hi}^{-1}(c)=b^*cb^*$,
    and hence
    $
      c\notin \supN(K,Q(L),P_{hi}) \parallel L
        = Q^{-1}(K-P_{hi}^{-1}P_{hi}(Q(L)-K)\Sigma_{hi}^*)\cap L
        = Q^{-1}(\{\eps,b\}) \cap L = \{\eps,a,b,ba\}
    $.
    Clearly, $supN(K,Q(L),P_{hi}) \| L \subseteq supN(K\|L, L, P)$. Therefore, OC does not guarantee that $\supN(K, Q(L), P_{hi}) \parallel L$ (at the high level) preserves $\supN(K \| L, L, P)$ (at the low level).
    \hfill$\diamond$
  \end{exmp}

  To guarantee the preservation of the supremal normal sublanguages, we modify the OC condition by fixing one of the low-level strings.

  \begin{defn}\label{defMOC}
    A prefix-closed language $L \subseteq \Sigma^*$ is {\em modified observation consistent} (MOC) with respect to projections $Q$, $P$, and $P_{hi}$ if for every string $s\in L$ and every string $t' \in Q(L)$ with $P_{hi}(Q(s)) = P_{hi}(t')$, there is a string $s' \in L$ such that $Q(s') = t'$ and $P(s) = P(s')$. \hfill$\Box$
  \end{defn}

\begin{figure}
    \centering
    \includegraphics[scale=.9]{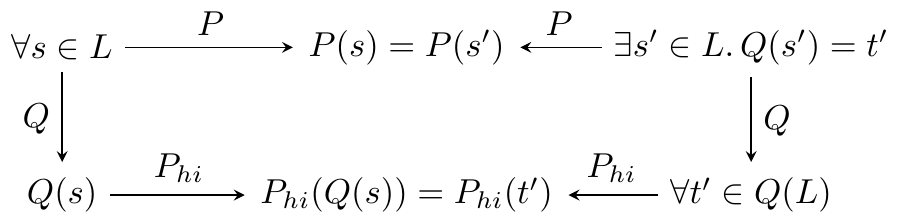}
    \caption{Illustration of the MOC condition.}
    \label{figMOC}
  \end{figure}

  Intuitively, every string that looks the same as a low-level string $s$ on the high level has a corresponding string that looks the same as $s$ on the low level, cf.~Figure~\ref{figMOC}.
  
  Notice that the language $L$ of Example~\ref{counterexample1} does not satisfy MOC, which may be verified directly or by Theorem~\ref{inclusionTHMmoc} below. Indeed, considering the strings $s=c\in L$ and $t'=bc\in Q(L)$, $P_{hi}(Q(c))=c=P_{hi}(bc)$, but for $s'\in Q^{-1}(bc)\cap L=\{bac\}$, $P(s)=c\neq ac=P(s')$. Hence the notions of OC and MOC are not equivalent.
  
  Comparing the notions of OC and MOC, MOC is stronger. 
  \begin{lem}\label{MOCstrongerOC}
    MOC implies OC.
  \end{lem}
  \begin{IEEEproof}
    If $L$ is MOC, then for any $t,t'\in Q(L)$ with $P_{hi}(t)=P_{hi}(t')$, there is $s\in L$ such that $t=Q(s)$. Hence, by MOC, there is $s'\in L$ such that $P(s)=P(s')$ and $Q(s')=t'$, which shows that $L$ is OC.
  \end{IEEEproof}

  The expression of MOC in terms of an inclusion of relational languages is similar to that of OC.
  
  \begin{thm}\label{inclusionTHMmoc}
    Let $L \subseteq \Sigma^*$ be a prefix-closed language, and let $\Sigma_o$ and $\Sigma_{hi}$ be the respective observation and high-level alphabets. Then $L$ is MOC w.r.t. $Q$, $P$, and $P_{hi}$ if and only if
    \[
      L\pp_{\Sigma_{hi}\cap \Sigma_o} Q(L) \subseteq Q_2\left(L\pp_{\Sigma_o} L\right)
    \]
    where $Q_2(a,b)=(a,Q(b))$ for every event pair $(a,b)$. 
  \end{thm}
  \begin{IEEEproof}
    We first show that if $L$ is MOC, then the inclusion holds. To this end, assume that $(s,t')\in L \pp_{\Sigma_{hi}\cap\Sigma_o} Q(L)$. By the definition of $\pp_{\Sigma_{hi}\cap\Sigma_o}$, we have that $s\in L$, $t' \in Q(L)$, and $P_{hi}(Q(s)) = P_{hi}(t')$. Since $L$ is MOC, there is $s' \in L$ such that $Q(s') = t'$ and $P(s) = P(s')$. However, $P(s)=P(s')$ implies that $(s,s')\in L \pp_{\Sigma_o} L$, and hence $Q(s')=t'$ implies that $(s,Q(s'))=(s,t')$, which shows the inclusion.
    
    On the other hand, we show that the inclusion implies that $L$ is MOC. For $s\in L$ and $t'\in Q(L)$ with $P_{hi}(Q(s))=P_{hi}(t')$, the definition of $\pp_{\Sigma_{hi}\cap\Sigma_o}$ implies that $(s,t')\in L \pp_{\Sigma_{hi}\cap\Sigma_o} Q(L)$. Since $L \pp_{\Sigma_{hi}\cap\Sigma_o} Q(L) \subseteq Q_2(L \pp_{\Sigma_o} L)$, we have that $(s,t')\in Q_2(L \pp_{\Sigma_o} L)$, and hence there is a pair $(s,s')\in L\pp_{\Sigma_o} L$ such that $(s,Q(s'))=(s,t')$. Since $(s,s')\in L\pp_{\Sigma_o} L$, strings $s$ and $s'$ belong to $L$ and coincide on $\Sigma_o$, i.\,e., $P(s)=P(s')$.
  \end{IEEEproof}

  We now show that the MOC condition guarantees that abstractions preserve normality of specification languages.

  \begin{lem}\label{lemma_supN}
    For a DFA $G$, let $L=L(G)$ and $L_m=L_m(G)$. If $L$ is MOC with respect to $Q$, $P$, and $P_{hi}$, then normality of $S\subseteq L_m$ with respect to $L$ and $P$ implies normality of $Q(S)$ with respect to $Q(L)$ and $P_{hi}$.
  \end{lem}
  \begin{IEEEproof}
     Since $Q(\overline{S}) \subseteq P_{hi}^{-1} P_{hi} (Q(\overline{S})) \cap Q(L)$, we need to show the opposite inclusion. Let $t'\in P_{hi}^{-1}P_{hi}(Q(\overline{S})) \cap Q(L)$. Then, there is $s\in \overline{S}$ such that $P_{hi}(Q(s))=P_{hi}(t')$. By MOC, there is $s'\in L$ such that $Q(s')=t'$ and $P(s)=P(s')$, i.\,e., $s'\in P^{-1}P(s)\cap L \subseteq P^{-1}P(\overline{S}) \cap L = \overline{S}$. Hence, $t'=Q(s')\in Q(\overline{S})$.
  \end{IEEEproof}

  We now show that, under the MOC condition, the high-level supervisor $S_{hi}$ realizing the supremal normal sublanguage of the specification $K$ with respect to the high-level plant $G_{hi}$ can be used to implement a nonblocking and maximally permissive low-level supervisor $S$ such that $L_m(S/G) = L_m(S_{hi}/G)$. In particular, the automaton realization $G_{S_{hi}}$ of the high-level supervisor $S_{hi}$ implements the low-level supervisor in the form of $L_m(S/G) = L_m(G_{S_{hi}} \parallel G)$.

  \begin{thm}\label{supN}
    For a nonblocking DFA $G$, let $L=L(G)$ and $L_m=L_m(G)$. If $L$ is MOC with respect to $Q$, $P$, and $P_{hi}$, then for every high-level specification $K\subseteq Q(L_m)$, 
    \[
      \supN\bigl(K\| L_m,L,P\bigr)
        = \supN(K,Q(L),P_{hi}) \parallel L_m
    \]
    whenever $\supN(K,Q(L),P_{hi})$ and $L_m$ are nonconflicting.
  \end{thm}
  \begin{IEEEproof}
    In order to prove the theorem, we need to show that $\supN\bigl(K\| L_m,L,P\bigr) = Q^{-1}(\supN(K,Q(L),P_{hi})) \cap L_m$. 
    
    Since $\supN\bigl(K\| L_m,L,P\bigr) \subseteq K\| L_m \subseteq L_m$, to prove that $\supN\bigl(K\| L_m,L,P\bigr) \subseteq Q^{-1}(\supN(K,Q(L),P_{hi})) \cap L_m$, we show that $\supN\bigl(K\| L_m,L,P\bigr) \subseteq Q^{-1}(\supN(K,Q(L),P_{hi}))$.
    However,
    $
      \supN\bigl(K\| L_m,L,P\bigr)
        \subseteq  Q^{-1}(\supN(K,Q(L),P_{hi}))
    $ holds if and only if the inclusion
    $
      Q(\supN\bigl(K\| L_m,L,P\bigr))
        \subseteq  \supN(K,Q(L),P_{hi})
    $ holds. Consider a language $S\subseteq K\| L_m$ normal w.r.t. $L$ and $P$. Then, by Lemma~\ref{lemma_supN}, $Q(S)\subseteq K\cap Q(L_m) = K$ is normal w.r.t. $Q(L)$ and $P_{hi}$, which shows that $Q(\supN\bigl(K\| L_m,L,P\bigr)) \subseteq \supN(K,Q(L),P_{hi})$, and finishes this part of the proof.

    To show $\supN(K,Q(L),P_{hi}) \| L_m \subseteq \supN\bigl(K\| L_m,L,P\bigr)$, we show that $\supN(K,Q(L),P_{hi}) \parallel L_m$ is normal w.r.t. $L$ and $P$. For brevity, we set $\supN=\supN(K,Q(L),P_{hi})$. Then
    \begin{align*}
      \overline{\supN \| L_m}
       & \subseteq P^{-1}P(\overline{\supN \| L_m}) \parallel \overline{L_m} \\
       & \subseteq P_{hi}^{-1}P_{hi}(\overline{\supN}) \parallel P^{-1}P(\overline{L_m}) \parallel \overline{L_m} \\
       & = P_{hi}^{-1}P_{hi}(\overline{\supN}) \parallel \overline{L_m} \\
       & = P_{hi}^{-1}P_{hi}(\overline{\supN}) \parallel  \overline{L_m}\parallel \overline{Q(L_m)}\\
       & = P_{hi}^{-1}P_{hi}(\overline{\supN}) \parallel  \overline{L_m}\parallel Q(\overline{L_m})\\
       & = P_{hi}^{-1}P_{hi}(\overline{\supN}) \parallel \overline{L_m}  \parallel Q(L)\\
       & = [P_{hi}^{-1}P_{hi}(\overline{\supN}) \cap Q(L)] \parallel \overline{L_m}\\
       & = \overline{\supN} \parallel \overline{L_m} 
       = \overline{\supN \| L_m}
    \end{align*}
    where the equality on the third line comes from the fact that $P^{-1}P(\overline{L_m}) \| \overline{L_m} = P^{-1}P(\overline{L_m}) \cap \overline{L_m} = \overline{L_m}$, because $\overline{L_m}\subseteq P^{-1}P(\overline{L_m})$, the last but one equality comes from normality of $\supN$, and the last equality as well as the second inclusion come from nonconflictingness of the languages. Therefore, $\supN(K,Q(L),P_{hi}) \parallel L_m \subseteq \supN(K\| L_m,L,P)$, as claimed.
  \end{IEEEproof}
  
  For prefix-closed specifications, we obtain the following.
  \begin{cor}\label{cor_supN}
    If $G$ is a DFA such that its language $L=L(G)$ is MOC with respect to $Q$, $P$, and $P_{hi}$, then
    \[
      \supN\bigl(K\| L,L,P\bigr)
        = \supN(K,Q(L),P_{hi}) \parallel L 
    \]
    for every prefix-closed high-level specification $K\subseteq Q(L)$.
    \hfill\IEEEQEDhere
  \end{cor}

  Recall that we only compute the high-level supremal normal sublanguage $\supN(K,Q(L),P_{hi})$, not the low-level supremal normal sublanguage $\supN\bigl(K\| L_m,L,P\bigr)$, achieving the closed-loop system $L_m(S_{hi}/G) = L_m(S/G)$.

\subsection{Railroad Controller}\label{railEx01}
\begin{figure}
    \centering
    \includegraphics[scale=.8]{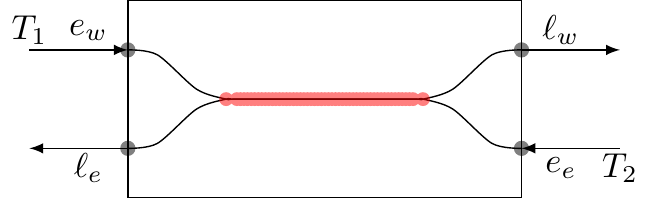}
    \caption{A bridge at the railroad where two tracks merge to a single track.}\label{figA1}
  \end{figure}

  To illustrate Theorem~\ref{supN}, we consider the synthesis of a bridge controller for a two-train railroad system motivated by Alur~\cite{Alur}.
  There are two circular tracks, one for trains traveling clockwise and the other for trains traveling counterclockwise. At one place, there is a bridge where the two tracks merge, cf. Figure~\ref{figA1} for an illustration.
  
\begin{figure}
    \centering
    \includegraphics[scale=.8]{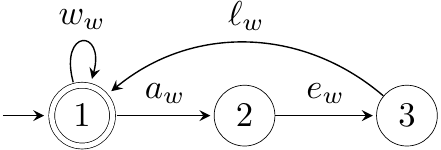}\quad
    \includegraphics[scale=.8]{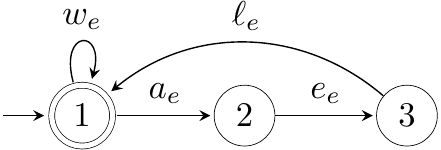}
    \caption{Generators $G_1$ and $G_2$ modeling the trains $T_1$ and $T_2$, resp.}\label{figGens}
  \end{figure}

  To control the access to the bridge, the trains communicate with a bridge controller, which we are constructing. In particular, there are two trains, the western train $T_1$ and the eastern train $T_2$. If the western train arrives at the bridge, it sends the arrive signal $a_w$. If the bridge controller accepts the signal, the train can enter the bridge ($e_w$); otherwise, it waits ($w_w$) and keeps sending the arrive signal $a_w$ until it is accepted. When leaving the bridge, the train sends the signal ($\ell_w$). The eastern train behaves the same way. The models of the two trains are depicted in Figure~\ref{figGens}.

\begin{figure}
    \centering
    \includegraphics[align=c,scale=.8]{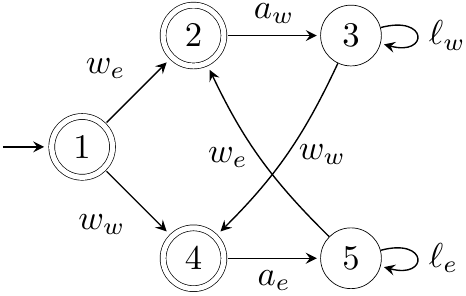}
    \quad
    \includegraphics[align=c,scale=.8]{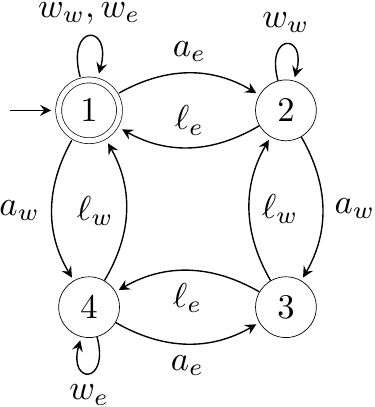}
    \caption{The specification $K$ (left) and the high-level plant $G_{hi}$ (right).}\label{figSpec1}\label{hiPlant}
  \end{figure}

  To construct the bridge controller, we consider the parallel composition $G_1 \| G_2$ of the train models as the plant, and post the following safety requirements on the supervisor.
  First, a train may enter the bridge only if its arrive signal is accepted. The arrive signal may be accepted if the other train waits (or is away from the bridge) and there is no train on the bridge. We define a specification that takes care not only about this requirement, but also ensures a kind of fairness, cf. Figure~\ref{figSpec1}.
  In particular, both trains wait before the arrive signal of one of the trains is accepted, and no train that wants to enter the bridge should wait for ever.

  To focus primarily on partial observation, we assume that all events are controllable. Suppose that the events $\ell_w$ and $\ell_e$ are unobservable, that is, $\Sigma_o=\{w_w, w_e, a_w, a_w, e_e, e_w\}$, and that the high-level alphabet is $\Sigma_{hi}=\{ w_w, w_e, a_w, a_w, \ell_2, \ell_w\}$.
  From the low-level plant $G=G_1 \| G_2$, which is obviously nonblocking because the plants share no events, we construct the high-level plant $G_{hi}$ as the projection of $G$ to the high-level alphabet $\Sigma_{hi}$, cf.~Figure~\ref{hiPlant}.\footnote{In this case, we may construct the high-level plant $G_{hi}$ as the parallel composition of the local high-level abstractions of the plants $G_1$ and $G_2$.}

  We can verify that the plant $G$ satisfies the MOC condition, and therefore, by Theorem~\ref{supN},
  $\supN(K\| L_m(G),L(G),P) = \supN(K,L(G_{hi}),P_{hi}) \| L_m(G)$.
 
\begin{figure}
    \centering
    \includegraphics[align=c,scale=.78]{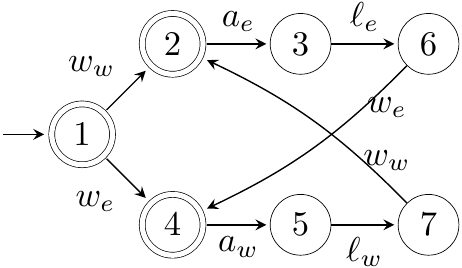}
    \quad
    \includegraphics[align=c,scale=.78]{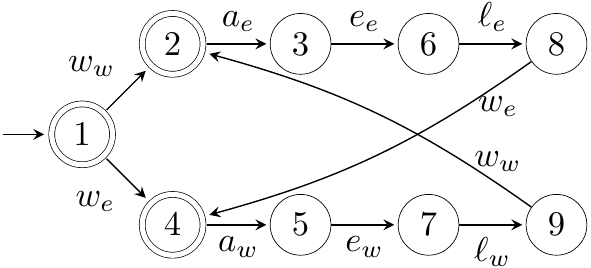}
    \caption{On the left, the language $\supN(K,L(G_{hi}),P_{hi})$, the high-level supervisor; and on the right, the language $\supN(K,L(G_{hi}),P_{hi}) \parallel L_m(G)$, the low-level supervisor.}\label{high-level-sup}
    \label{supN_KparLm}
  \end{figure}

  Finally, we construct the high-level supervisor $S_{hi}$ realizing the language $\supN(K,L(G_{hi}),P_{hi})$, depicted in Figure~\ref{high-level-sup}. Theorem~\ref{supN} now implies that the closed-loop system $L_m(S_{hi}/G)=\supN(K,L(G_{hi}),P_{hi}) \| L_m(G)$ coincides with the closed-loop $L_m(S/G)=\supN(K\| L_m(G),L(G),P)$ for a nonblocking and maximally permissive low-level supervisor $S$.
  The high-level and low-level supervisors are compared in Figure~\ref{supN_KparLm}.

\section{Decidability and Complexity of (M)OC}
\label{section6}\label{decMOC}
  A {\em decision problem\/} is a yes-no question. A decision problem is {\em decidable\/} if there is an algorithm solving it. Complexity theory classifies decidable problems into classes based on time or space an algorithm needs to solve the problem. The complexity class we consider is \PSpace, denoting all problems solvable by a deterministic polynomial-space algorithm. A decision problem is \PSpace-hard if every problem from \PSpace can be reduced to it by a polynomial-time algorithm. If the problem belongs to \PSpace, it is \PSpace-complete. It is a longstanding open problem whether \PSpace-complete problems can be solved in polynomial time.

  We show that verifying (M)OC is \PSpace-hard for systems modeled by finite automata. 
  
  \begin{thm}\label{thm5}
    Verifying (M)OC for DFAs is \PSpace-hard.
  \end{thm}
  \begin{IEEEproof}
    We reduce the \PSpace-complete problem of deciding universality for NFAs with all states marked~\cite{KaoRS09}. The problem asks whether, given an NFA $A$ over $\Sigma$ with all states marked, its language $L(A)=\Sigma^*$. These NFAs recognize prefix-closed languages, i.\,e., $L(A)=L_m(A)$.
   
    Let $A=(Q,\Sigma,\delta,q_0,Q)$ be an NFA with all states marked. We first modify $A$ to a DFA by adding additional transitions. 
    Namely, for every transition $p\xrightarrow{a} q$ of $A$, we add a new marked state $x_{p,a,q}$ and a new event $a_{p,a,q}$, and replace the transition $p\xrightarrow{a} q$ with two transitions $p\xrightarrow{a_{p,a,q}} x_{p,a,q} \xrightarrow{a} q$. We denote the resulting automaton by $A'$ and the set of newly added events by $\Sigma'$. Obviously, $A'$ is deterministic with all states marked, i.\,e., $L(A')=L_m(A')$. Further, a string $w$ belongs to $L(A)$ if and only if there is a path in $A'$ where the events on odd positions belong to $\Sigma'$, and the sequence of events on even positions forms $w$; i.\,e., $a_1\ldots a_n\in L(A)$ if and only if $\Sigma'a_1\Sigma'\cdots\Sigma'a_n \cap L(A')\neq\emptyset$.

    In the second step, we construct a DFA $B$ such that 
    \[
      L_m(B) = @\#L(A') \cup @(\Sigma'\Sigma)^* \cup \#(\Sigma'\Sigma)^* \cup L(A')
    \]
    where $@$ and $\#$ are new events. It is not difficult to construct $B$ from $A'$ in polynomial time. In particular, to ensure that $L_m(B)$ accepts everything from $L(A')$, we add the transition $n_1 \xrightarrow{a} q$, where $n_1$ is the initial state of $B$, whenever there is a transition $q_0 \xrightarrow{a} q$ for some state $q$ in $A'$. Figure~\ref{fig0} illustrates the construction for two transitions from $q_0$ under $a$ and $b$.

    \begin{figure}
      \centering
      \includegraphics[scale=.8]{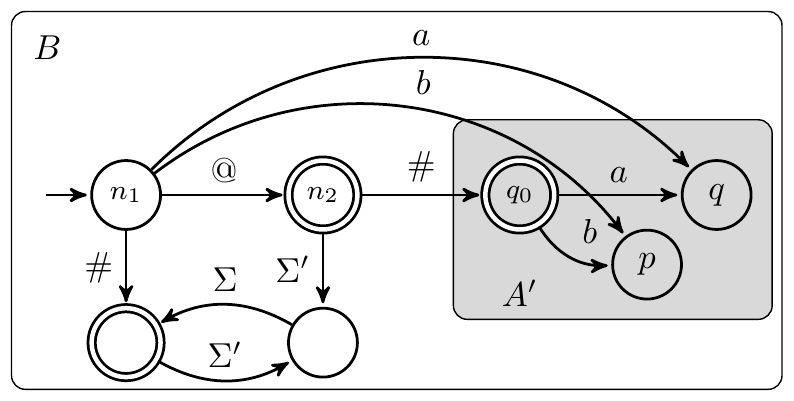}
      \caption{Construction of the DFA $B$ from the DFA $A'$.}
      \label{fig0}
    \end{figure}
    
    Let the abstraction $Q$ remove $\Sigma'\cup\{@\}$, and the observation $P$ remove $\Sigma'\cup\{\#\}$, i.\,e., $\Sigma_{hi}=\Sigma\cup\{\#\}$ and $\Sigma_o=\Sigma\cup\{@\}$. Then $Q(L(B))=\Sigma^* \cup \#\Sigma^*$. We now show that $L(B)=\overline{L_m(B)}$ is (M)OC if and only if $A$ is universal.

    Assume that $A$ is universal, i.\,e., $L(A)=\Sigma^*$. Let $s\in L(B)$ and $Q(s)\neq t'\in Q(L(B))$ with $P_{hi}(Q(s)) = P_{hi}(t')$. By the construction of $B$, we have two cases:
    
    First, $Q(s)\in\Sigma^*$ and $t'=\#Q(s)$ for $s\in \overline{@(\Sigma'\Sigma)^*} \cup L(A')$.
    If $s\in \overline{@(\Sigma'\Sigma)^*}$, then $s = @ x_1 s_1 x_2 \cdots x_{k} s_{k} x_{k+1}$, where $x_1,\ldots,x_k\in \Sigma'$, $x_{k+1}\in \Sigma' \cup \{\eps\}$, and $s_1\cdots s_k \in L(A) = \Sigma^*$. Then, taking $s' = @ \# x_1 s_1 x_2 \cdots x_{k} s_{k} x_{k+1} \in @ \# \Sigma' s_1 \Sigma' \cdots \Sigma' s_k (\Sigma' \cup \{\eps\}) \subseteq @ \# L(A') \subseteq L(B)$ is such that $Q(s') = \# Q(s) = t'$ and $P(s) = @ s_1 \cdots s_k = P(s')$.
    If $s\in L(A')$, then $s = x_1 s_1 x_2 \cdots x_k s_k x_{k+1}$, where $x_1, \ldots, x_k \in \Sigma'$, $x_{k+1} \in \Sigma' \cup \{\eps\}$, and $s_1\cdots s_k \in L(A)$. Then, $s' = \# x_1 s_1 x_2 \cdots x_k s_k x_{k+1} \in \# \Sigma' s_1 \Sigma' \cdots \Sigma' s_k (\Sigma' \cup \{\eps\}) \subseteq L(B)$ satisfies $Q(s')=\#Q(s)=t'$ and $P(s)=P(s')$.

    Second, $t'\in\Sigma^*$ and $Q(s)=\#t'$ for $s \in @ \# L(A') \cup \overline{\#(\Sigma'\Sigma)^*}$.
    If $s\in @ \# L(A')$, then we have that $s = @ \# x_1 t'_1 x_2 \cdots x_k t'_k x_{k+1}$, where $x_1,\ldots,x_k \in \Sigma'$, $x_{k+1}\in \Sigma'\cup\{\eps\}$, and $t'=t_1'\cdots t_k'$. Therefore, taking $s' = @ x_1 t'_1 x_2 \cdots x_k t'_k x_{k+1} \in L(B)$ satisfies $Q(s')=t'$ and $P(s)=@t'=P(s')$.
    If $s\in \overline{\#(\Sigma'\Sigma)^*}$, then we have that $s = \# x_1 t'_1 x_2 \cdots x_k t'_k x_{k+1}$, where $x_1,\ldots,x_k \in \Sigma'$, $x_{k+1}\in \Sigma'\cup \{\eps\}$, and $t'=t_1'\cdots t_k'$. Then, $s' = x_1 t'_1 x_2 \cdots x_k t'_k x_{k+1} \in L(A') \subseteq L(B)$ is such that $Q(s')=t'$ and $P(s)=t'=P(s')$.
    Thus, $B$ is MOC, and hence OC by Lemma~\ref{MOCstrongerOC}.

    On the other hand, if $A$ is not universal, then there exists $w=a_1a_2\cdots a_k\notin L(A)$. Then, for any $x_1,x_2\ldots,x_k \in \Sigma'$, we have that $@w' = @x_1a_1x_2a_2\cdots x_ka_k \in @(\Sigma'\Sigma)^*$ and that $\#w' = \#x_1a_1x_2a_2\cdots x_ka_k \in \#(\Sigma'\Sigma)^*$. Therefore, the strings $@w',\#w' \in L(B)$, $Q(\#w')=\#w\in Q(L(B))$, and $P_{hi}(Q(@w'))=P_{hi}(w)=w=P_{hi}(\#w)$. We now show that there are no $s,s'\in L(B)$ such that $Q(s)=Q(@w')=w$, $Q(s')=\#w$, and $P(s)=P(s')$, which means that $L(B)$ is not OC. By Lemma~\ref{MOCstrongerOC}, $L(B)$ is then neither MOC. 
    To this end, because $w=a_1 a_2\cdots a_k \notin L(A)$, we may observe that 
    $
      Q^{-1}(w)\cap L(B) \subseteq @\Sigma'a_1 \Sigma' a_2 \cdots \Sigma'a_{k}(\Sigma' \cup \{\eps\})
    $
    and 
    $
      Q^{-1}(\#w) \cap L(B) \subseteq \#\Sigma' a_1 \Sigma' a_2 \cdots \Sigma' a_{k}(\Sigma' \cup \{\eps\})\,.
    $
    But then, for any $s\in Q^{-1}(w)\cap L(B)$ and $s'\in Q^{-1}(\#w)\cap L(B)$, we obtain that
    $
      P(s) = P(@\Sigma'a_1\cdots\Sigma'a_{k}(\Sigma' \cup \{\eps\}))= @w \neq w 
      = P(\#\Sigma'a_1 \cdots \Sigma' a_{k}(\Sigma' \cup \{\eps\})) = P(s')
    $,
    which completes the proof.
  \end{IEEEproof}

  In Appendix~\ref{appendixE}, we show that if the conditions are decidable, then the regularity of the system is essential. Namely, for non-regular systems modeled by one-turn deterministic pushdown automata that are slightly more expressive than finite automata, the conditions are undecidable.

\section{Practical Conditions}\label{section7}\label{sec:Practical Conditions}
  As shown above, verifying (M)OC is a hard and maybe even undecidable problem. It is therefore reasonable to consider stronger and easily-verifiable conditions. One of such conditions often considered in the literature is that all observable events are high-level events, i.\,e., $\Sigma_o\subseteq \Sigma_{hi}$.
  In this section, we consider two such conditions:
  \begin{enumerate}
    \item $\Sigma_o\subseteq \Sigma_{hi}$, i.\,e., all observable events are high-level, and
    \item $\Sigma_{hi}\subseteq \Sigma_{o}$, i.\,e., all high-level events are observable.
  \end{enumerate}
  
  We show that both these conditions imply (M)OC.

  \begin{thm}\label{cond1}\label{cond2}
    If $G$ is a DES over $\Sigma$ satisfying either $\Sigma_o\subseteq \Sigma_{hi}$ or $\Sigma_{hi}\subseteq \Sigma_{o}$, then $L=L(G)$ is (M)OC.
  \end{thm}
  \begin{IEEEproof}
    If $\Sigma_o\subseteq \Sigma_{hi}$, then $P=P_{hi}Q$, because $Q_o$ is an identity, cf. Figure~\ref{projections}. Let $s\in L$ and $t'\in Q(L)$ be strings such that $P_{hi}(Q(s))=P_{hi}(t')$, and consider any $s'\in L$ such that $Q(s')=t'$. Such a string $s'$ exists, because $t'\in Q(L)$, and hence the following equalities $P(s) = P_{hi}(Q(s)) = P_{hi}(t') = P_{hi}(Q(s')) = P(s')$ prove that $L$ satisfies MOC. 
    If $\Sigma_{hi}\subseteq \Sigma_{o}$, then $P_{hi}$ is an identity. Hence, for any $s\in L$ and $t'\in Q(L)$ with $P_{hi}(Q(s)) = P_{hi}(t')$, we have $Q(s) = P_{hi}(Q(s)) = P_{hi}(t') = t'$, i.\,e., we can chose $s'=s$ to satisfy MOC. 
    That $L$ also satisfies OC then follows from Lemma~\ref{MOCstrongerOC}.
  \end{IEEEproof}
  
  Theorem~\ref{cond2} is widely used in  Section~\ref{section8}, which illustrates its applicability.

  The reader may wonder whether the MOC condition is not equivalent to $\Sigma_o \subseteq \Sigma_{hi}$ or $\Sigma_{hi} \subseteq \Sigma_o$. It is not the case. First, deciding $\Sigma_o \subseteq \Sigma_{hi}$ and $\Sigma_{hi} \subseteq \Sigma_o$ is computationally a simple task, whereas deciding MOC is \PSpace-hard. Second, Boutin et al.~\cite[Example~1]{cdc-ecc2011} show that the converse of Theorem~\ref{cond1} does not hold for OC. In fact, the system of this example also satisfies MOC, as the reader can verify using, e.g., Theorem~\ref{inclusionTHMmoc}, showing thus that the converse of Theorem~\ref{cond1} neither holds for MOC.

  As an immediate consequence, Theorem~\ref{supN} strengthens the claim of Komenda and Masopust~\cite{KM10} showing that for any prefix-closed languages $L\subseteq \Sigma^*$ and $K\subseteq Q(L)$, if $\Sigma_o\subseteq \Sigma_{hi}$, then $\supN(K,Q(L),P_{hi})\parallel L = \supN(K\| L,L,P)$. 

  Another practically interesting case is to find a condition weaker than both (i) and (ii) that would imply (M)OC. The existence of such a condition is, however, an open problem.

  \begin{remark}
    The conditions can also be used if $\Sigma_o$ and $\Sigma_{hi}$ are incomparable. Namely, we (i) compute the high-level supervisor $S_{hi}$, and (ii) take the system $G$ with observable events $\Sigma_o'=\Sigma_o \cup \Sigma_{hi}$, which implies MOC, and construct a high-level supervisor $S_{hi}'$. If $L_m(S_{hi}/G) = L_m(S_{hi}'/G)$, then $S_{hi}$ running in parallel with the low-level plant realizes a low-level supervisor; otherwise, we need to verify MOC.
  \end{remark}

\section{Case Study}\label{section8}
  To evaluate presented conditions and results on an industrial example, we consider the model and specification of a patient table of an MRI scanner as presented by Theunissen~\cite{theunissen}. The plant consists of four components
  \[
    \text{VAxis}  \parallel  \text{HAxis} \parallel  \text{HVNormal}  \parallel \text{UI}
  \]
  where each component is again a composition of other components. However, we do not go into further details and consider these components as the four low-level subsystems forming the global low-level plant. Similarly, the specification consists of the corresponding components
  \[
    \text{VReq}  \parallel  \text{HReq}  \parallel  \text{HVReq}  \parallel \text{UIReq}
  \]
  which do not exactly fit the four subsystems of the low-level plant. The aim is to construct four high-level supervisors, one for each specification and the corresponding plant, see the details below. For the computations, we used the C++ library \texttt{libFAUDES} version 2.29b~\cite{libfaudes}. The computations were performed on an Inter-Core i7 processor laptop with 15 GB memory, running Ubuntu 20.04.

  In the model of Theunissen~\cite{theunissen}, all events are observable. Therefore, the idea of our approach is to consider the events that appear in a specification as observable, and all the other events as unobservable. Since we are currently unable to algorithmically verify whether the MOC condition is satisfied or not, and the models are too large for a by-hand verification, we ensure MOC by using the stronger conditions of Section~\ref{section7}. Namely, we proceed as follows:
  \begin{outline}
    \1[1)] For every $K\in\{\text{VReq},\text{HReq},\text{HVReq},\text{UIReq}\}$,
    we take all plants from $\{\text{VAxis},\text{HAxis},$ $\text{HVNormal},\text{UI}\}$ that share an event with $K$, and define the low-level plant, denoted by $G_{low}$, as their parallel composition.

    \1[2)] We define $\Gamma_o$ as the set of all events occurring in $K$, and we set $\Delta_{hi} = \Gamma_o$; that is, all events of $K$ are set to be high level and observable.

    \1[3)] Existing results on controllability of hierarchical supervisory control extend the high-level alphabet $\Delta_{hi}$ to $\Gamma_{hi}$ to guarantee that the low-level plant $G_{low}$ is an $L_m(G_{low})$-observer and LCC. To do this, we use the \texttt{libFAUDES} function \texttt{NaturalObserverLcc}.
      \2 We now have $\Gamma_o \subseteq \Gamma_{hi}$, and Theorem~\ref{cond1} implies that the low-level plant $G_{low}$ satisfies MOC w.r.t. $\Gamma_o$ and $\Gamma_{hi}$, which makes Theorem~\ref{supN} applicable.

    \1[4)] For a moment, we define $\Sigma_o = \Sigma_{hi} = \Gamma_{hi}$, and compute the high-level plant, $G_{hi}'$, and the nonblocking and maximally permissive high-level supervisor, $S_{hi}'$, realizing the supremal controllable and normal sublanguage of the high-level specification w.r.t. the high-level plant $G_{hi}'$.
      \2 The high-level specification is $(P_{\Gamma_o}^{\Gamma_{hi}})^{-1}(K)$, and is obtained by lifting $K$ from the alphabet $\Gamma_o$ to $\Gamma_{hi}$ by the inverse of the projection $P_{\Gamma_o}^{\Gamma_{hi}}\colon \Gamma_{hi}^* \to \Gamma_o^*$.
      \2 This gives us a referential closed-loop $L_m(S_{hi}'/G)$.

    \1[5)] We now try to find a high-level supervisor $S_{hi}$ represented by a smaller automaton than the automaton representing $S_{hi}'$. To this end, we search for alphabets $\Sigma_o$ and $\Sigma_{hi}$, such that $\Gamma_o \subseteq \Sigma_o \subseteq \Sigma_{hi} \subseteq \Gamma_{hi}$ or $\Gamma_o \subseteq \Sigma_{hi} \subseteq \Sigma_o \subseteq \Gamma_{hi}$, for which the nonblocking and maximally permissive high-level supervisor $S_{hi}$, constructed for the high-level plant, $G_{hi}$, w.r.t. the alphabets $\Sigma_o$ and $\Sigma_{hi}$, satisfies $L_m(S_{hi}/G) = L_m(S_{hi}'/G)$.
      \2 We first try the alphabets $\Sigma_{hi} = \Sigma_o = \Gamma_o$.
      \2 If it fails, we set $\Sigma_{o} = \Gamma_{hi}$, which is a sufficient observation to achieve the referential closed-loop system $L_m(S_{hi}'/G)$, and we search for a suitable alphabet $\Sigma_{hi}$, such that $\Gamma_o \subseteq \Sigma_{hi}\subseteq \Sigma_o$, for which $L_m(S_{hi}/G) = L_m(S_{hi}'/G)$.
        \3 $\Sigma_{hi} \subseteq \Sigma_{o}$ still makes Theorem~\ref{supN} applicable.
      \2 Otherwise, we take $S_{hi}=S_{hi}'$.
  \end{outline}

  Automata representations of all considered supervisors are computed using the \texttt{libFAUDES} function \texttt{SupConNormNB}, which implements the standard algorithm for the computation of the supremal controllable and normal sublanguage. All the automata are further minimized w.r.t. the number of states.

  The reader may notice that, in Step~5, we do not consider all the possible choices for the alphabets $\Sigma_o$ and $\Sigma_{hi}$, because it would be computationally demanding. For the same reason, we do not even consider all the choices for the alphabet $\Sigma_{hi}$, such that $\Gamma_o \subseteq \Sigma_{hi}\subseteq \Sigma_o$. In particular, we do not consider the cases where the alphabets are incomparable, which opens the door for further improvements to be investigated.

\subsection{Results for the Four Specifications}
  We summarize the results for each of the four specifications in the following tables. For comparison, we include the statistics of the automaton representation of the nonblocking and maximally permissive low-level supervisor, $S_{low}$, realizing the supremal controllable and normal sublanguage of the low-level specification, obtained from $K$ by the inverse projection to the low-level alphabet, w.r.t. the low-level plant $G_{low}$.

  The specification VReq consists of nine events shared only with the plant VAxis, i.\,e., $G_{low}=\text{VAxis}$. These events form the set $\Gamma_o$. The computation was successful for the choice of $\Sigma_{hi}=\Sigma_o=\Gamma_o$. Table~\ref{tabVAxis} shows the results.

  \begin{table}[ht]
    \ra{1.1}
    \centering
    \caption{}\label{tabVAxis}
    \begin{tabular}{lrrrrr}
      \toprule
              & VReq & VAxis & $G_{hi}$ & $S_{hi}$ & $S_{low}$\\
                \midrule
      States  & 12 & 15 & 15 & 11 & 15 \\
      Trans.  & 44 & 50 & 44 & 22 & 36 \\
      Events  &  9 & 11 & 9  &  9 & 11 \\
      \bottomrule
    \end{tabular}
  \end{table}

  The specification HReq consists of 19 events, forming the set $\Gamma_o$, occurring only in HAxis, and hence $G_{low}=\text{HAxis}$. Again, the computation was successful for $\Sigma_{hi}=\Sigma_o=\Gamma_o$, and the results are summarized in Table~\ref{tabHAxis}.

  \begin{table}[h]
    \ra{1.1}
    \centering
    \caption{}\label{tabHAxis}
    \begin{tabular}{lrrrrr}
      \toprule
              & HReq & HAxis & $G_{hi}$ & $S_{hi}$ & $S_{low}$\\
                \midrule
      States  & 112 & 128  & 128 & 80  & 80 \\
      Trans.  & 736 & 1002 & 986 & 312 & 320 \\
      Events  &  19 & 20   &  19 &  19 & 20 \\
      \bottomrule
    \end{tabular}
  \end{table}

  The specification HVReq consists of ten events, forming the set $\Gamma_o$, occurring in VAxis, HAxis, and HVNormal, and hence $G_{low} = \text{VAxis} \parallel \text{HAxis} \parallel \text{HVNormal}$. However, the choice of $\Sigma_{hi}=\Sigma_o=\Gamma_o$ fails, and therefore we set $\Sigma_{o} = \Gamma_o'$, which has 26 events, and find $\Sigma_{hi}\subseteq \Sigma_o$ with 16 events. The results are summarized in Table~\ref{tabHVAxis2}.
  \begin{table}[h]
    \ra{1.1}
    \centering
    \caption{}\label{tabHVAxis2}
    \begin{tabular}{lrrrrr}
      \toprule
              & HVReq & $G_{low}$ & $G_{hi}$ & $S_{hi}$ & $S_{low}$\\
                \midrule
      States  & 7  & 1920  & 320  & 381  & 2064 \\
      Trans.  & 35 & 23350 & 2638 & 2216 & 20120 \\
      Events  & 10 & 32    &  16  &  16  & 32 \\
      \bottomrule
    \end{tabular}
  \end{table}

  Finally, the specification UIReq has 21 events, forming the set $\Gamma_o$, shared with VAxis, HAxis, HVNormal, and UI, and hence $G_{low} = \text{VAxis} \parallel \text{HAxis} \parallel \text{HVNormal} \parallel \text{UI}$. The choice of $\Sigma_{hi}=\Sigma_o=\Gamma_o$ was successful, cf. Table~\ref{tabUIReq}.

  \begin{table}[h]
    \ra{1.1}
    \centering
    \caption{}\label{tabUIReq}
    \begin{tabular}{lrrrrr}
      \toprule
             & UIReq & $G_{low}$ & $G_{hi}$ & $S_{hi}$ & $S_{low}$\\
                \midrule
      States & 256  & 3840  & 64   & 3296  & 211200  \\
      Trans. & 2336 & 75500 & 1080 & 28936 & 2751680 \\
      Events & 21   & 41    &  21  &  21  & 41 \\
      \bottomrule
    \end{tabular}
  \end{table}

  We should point out that all the reference supervisors $S_{hi}'$ computed in Step~4 of the construction were also smaller in all the considered statistics than the low-level supervisor $S_{low}$.

  To summarize, we computed four nonblocking and maximally permissive high-level supervisors with altogether $3768$ states and $31486$ transitions, which in parallel with the low-level plants exactly achieve the behavior of the nonblocking and maximally permissive low-level supervisor. The computed supervisors are nonblocking and nonconflicting.
  For comparison, the four low-level supervisors have altogether $213359$ states and $2772156$ transitions, while the monolithic low-level supervisor, denoted by $S_{mono}$, has 68672 states and 616000 transitions, see Table~\ref{tabSum}.

  \begin{table}[h]
    \ra{1.1}
    \centering
    \caption{}\label{tabSum}
    \begin{tabular}{lrrrrr}
      \toprule
             & $4 \times S_{hi}$ & $4 \times S_{low}$ & $S_{mono}$\\
                \midrule
      States & 3768  & 213359  & 68672\\
      Trans. & 31486 & 2772156 & 616000\\
      \bottomrule
    \end{tabular}
  \end{table}

  In fact, the behaviors of the above-discussed low-level supervisors computed under partial observation coincide with the corresponding supremal controllable sublanguages computed under complete observation. In other words, the considered partial observation did not restrict the supervisors compared with the completely observed systems.
  
\subsection{The Worst Experimental Time Complexity}
  From the experimental time-complexity viewpoint, the most time-consuming were the computations for the specification UIReq. Namely, it took 11 seconds to compute the high-level plant $G_{hi}$ and one second to compute the high-level supervisor $S_{hi}$. (For comparison, in the other cases, the computations took time in the order of milliseconds.)
  
  On the other hand, the computation of the low-level supervisor for UIReq ran out of memory after one hour and 14 minutes.
  In contrast, computing this supervisor as the parallel composition of the high-level supervisor $S_{hi}$ with the low-level plant $G_{low}$ according to Theorem~\ref{supN} took only 14 seconds.
  
  Finally, as already pointed out above, the low-level supremal controllable and normal sublanguage of the specification UIReq computed under the considered partial observation coincides with the low-level supremal controllable sublanguage when considering all events observable. In comparison, the low-level supremal controllable sublanguage was computed in a few seconds.

  \begin{remark}
    The reader may notice that although our low-level plants are partially observed, the high-level abstractions are completely observed. Therefore, our choice of abstractions satisfying the MOC condition reduces the problem of supervisory control synthesis under partial observation to the problem of supervisory control synthesis under complete observation, where the latter is computationally significantly easier; see also the discussion above.
  \end{remark}

\section{Conclusion}\label{section9}
  We suggested a new sufficient condition, MOC, that guarantees maximal permissiveness of supervisors in hierarchical supervisory control under partial observation. Since decidability and complexity of its verification for systems modeled by finite automata is an open problem, we showed that the finite-automata models are essential for potential decidability, because the problem is undecidable for slightly more expressive models of one-turn deterministic pushdown automata. We further discussed several relevant conditions that ensure MOC.

  In an upcoming work, we will discuss applications of the results in modular supervisory control under partial observation.
  In the future, we will explore the decidability of (M)OC and the constructions of high-level relatively-observable supervisors. Our plan is to further extend the hierarchical approach to modular supervisory control of networked DES to account for delays and losses in observation channels.

\appendices
\section{Undecidability of (M)OC in Non-Regular Systems}\label{appendixE}
  We now show that for a slightly more expressive model than DFAs, verifying (M)OC is undecidable.
    
  A {\em pushdown automaton\/} is a septuple $\mathcal{M} = (Q, \Sigma, \Gamma, \delta, q_0,$ $Z_0, F)$, where $Q$ is a finite set of states, $\Sigma$ is an alphabet, $\Gamma$ is a pushdown alphabet, $\delta\colon Q\times (\Sigma\cup\{\eps\})\times \Gamma \to 2^{Q\times\Gamma^*}$ is the transition function, $q_0\in Q$ is the initial state, $Z_0\in\Gamma$ is the initial pushdown symbol, and $F\subseteq Q$ is the set of accepting states.
  A {\em configuration} of $\mathcal{M}$ is a triplet $(q,w,\gamma)$, where $q\in Q$ is the current state, $w\in \Sigma^*$ is the unread part of the input, and $\gamma\in \Gamma^*$ is the current content of the pushdown (the leftmost symbol of $\gamma$ represents the top pushdown symbol). For $p,q\in Q$, $a\in\Sigma\cup\{\eps\}$, $w\in\Sigma^*$, $\gamma,\beta\in\Gamma^*$, $Z\in\Gamma$, and $(p,\beta)\in\delta(q,a,Z)$, $\mathcal{M}$ makes a move from $(q,aw,Z\gamma)$ to $(p,w,\beta\gamma)$, denoted by $(q,aw,Z\gamma) \vdash_\mathcal{M} (p,w,\beta\gamma)$. The reflexive and transitive closure of the relation $\vdash_\mathcal{M}$ is denoted by $\vdash^*_\mathcal{M}$. The {\em language accepted} by $\mathcal{M}$ is the set $L(\mathcal{M})=\{w\in\Sigma^* \mid (q_0,w,Z_0) \vdash^*_\mathcal{M} (q,\eps,\gamma) \textrm{ for some } q\in F \textrm{ and } \gamma\in\Gamma^*\}$.
  A pushdown automaton $\mathcal{M}$ is {\em deterministic} (DPDA) if 
    (1) $|\delta(q,a,Z)|\le 1$, for all $a\in\Sigma\cup\{\eps\}$, $q\in Q$, and $Z\in\Gamma$, and 
    (2) for all $q\in Q$ and $Z\in\Gamma$, if $\delta(q,\eps,Z)\neq\emptyset$, then $\delta(q,a,Z)=\emptyset$, for all $a\in\Sigma$.
  During the computation of a DPDA, the height of its pushdown increases and decreases. The situation where the pushdown changes from the increasing phase to the decreasing phase, or vice versa, is a {\rm turn}~\cite{GS66}.
  A language is {\em linear deterministic context-free\/} if it is accepted by a one-turn DPDA (1-DPDA), i.\,e., by a DPDA whose pushdown content first only increases and then only decreases.
  
  Linear deterministic context-free languages are slightly more expressive than regular languages. Intuitively, linear deterministic context-free languages are a generalization of languages of palindromes~\cite{berstel}. Perhaps a better argument expressing the closeness of linear deterministic context-free languages to regular languages is provided by Rosenberg~\cite{Rosenberg67}, who relates linear languages to 2-tape finite automata.

  \begin{thm}\label{undec1}
    Verifying (M)OC for 1-DPDAs is undecidable. 
  \end{thm}
  \begin{IEEEproof}
    We prove the theorem by reduction of the Post's Correspondence Problem (PCP)~\cite{post} to (M)OC. The PCP asks whether, given two finite lists $A=(w_1,\ldots,w_n)$ and $B=(u_1,\ldots,u_n)$ of $n$ strings over $\Sigma$, there is a sequence of indices $i_1 i_2 \cdots i_k$, for some $k\ge 1$, such that $w_{i_1} w_{i_2} \cdots w_{i_k} = u_{i_1} u_{i_2} \cdots u_{i_k}$. To eliminate trivial cases, we may assume that $w_i\neq u_i$, for $i=1,\ldots, n$. 
    We denote by $E=\{1,\ldots,n\}$ a new alphabet, such that $E\cap\Sigma=\emptyset$, to represent the indices of words in the lists $A$ and $B$, and by $w^R$ the mirror image of $w\in\Sigma^*$, e.g., $(abc)^R = cba$. Now, we define the language 
    \begin{multline*}
      L=
      \{i_1i_2 \cdots i_m \$ w_{i_m}^R \cdots w_{i_2}^Rw_{i_1}^R@ \mid m\ge 1\} \\
      \cup
      \{i_1i_2 \cdots i_m \$' u_{i_m}^R \cdots u_{i_2}^Ru_{i_1}^R\# \mid m\ge 1\}
    \end{multline*}
    where $\$$, $@$, $\$'$, and $\#$ are new events.
    
    Languages $L$ and $\overline{L}$ are accepted by a 1-DPDA: For a given input string, the 1-DPDA initially reads the events from $E$ and pushes them to the pushdown. Then it reads $\$$ or $\$'$, which specifies whether the rest of the input consists of the mirrored strings from $A$ or from $B$, respectively. Next, it pops events one by one from the pushdown, say $i$, and tries to read the corresponding mirrored string $w_{i}^R$, resp. $u_{i}^R$, from the input. To recognize $L$, the 1-DPDA accepts an input string if it successfully empties the pushdown and reads the whole input string; otherwise, it rejects. To recognize $\overline{L}$, the 1-DPDA accepts if it successfully reads the whole input.
    
    Let 
      $\Sigma_{hi}=\Sigma\cup E\cup\{@,\#\}$ 
    and 
      $\Sigma_o=\Sigma\cup E \cup\{\$,\$'\}$. 
    We prove that the instance has a solution iff $\overline{L}$ is not (M)OC. 
    
    Assume that $i_1,\ldots,i_k$ is a solution of the instance of PCP. We consider the two strings $r = i_1\ldots i_k \$ w_{i_k}^R\ldots w_{i_1}^R@$ and $t'=i_1\ldots i_k u_{i_k}^R\ldots u_{i_1}^R\#$, and show that $t=Q(r)=i_1\ldots i_k w_{i_k}^R\ldots w_{i_1}^R@$ and $t'$ violate OC, which, by Lemma~\ref{MOCstrongerOC}, violates MOC as well. To this end, notice first that $P_{hi}(t)=P_{hi}(Q(r)) = i_1\ldots i_k w_{i_k}^R\ldots w_{i_1}^R = i_1\ldots i_k u_{i_k}^R\ldots u_{i_1}^R = P_{hi}(t')$, and hence we need to show that for any $s,s'\in \overline{L}$ with $Q(s)=t$ and $Q(s')=t'$, the observations of $s$ and $s'$ are different, i.\,e., $P(s)\neq P(s')$. However, $s\in Q^{-1}(t)\cap L = \{i_1 \cdots i_k \$ w_{i_k}^R \cdots w_{i_1}^R@\}$ contains the symbol $\$$, whereas $s'\in Q^{-1}(t')\cap L = \{i_1 \cdots i_k \$' u_{i_k}^R \cdots u_{i_1}^R\#\}$ contains $\$'$ but no $\$$, and hence $P(s')\neq P(s)$ and $L$ is not (M)OC.
    
    On the other hand, we assume that the instance of PCP has no solution. In this case, we consider any two strings $s\in \overline{L}$ and $t'\in Q(\overline{L})$ satisfying $P_{hi}(Q(s))=P_{hi}(t')$. 
    Then, if $Q(s)=t'$, we may simply take $s'=s$ to obtain that $Q(s)=Q(s')=t'$ and $P(s)=P(s')$, as required by MOC. Otherwise, if $Q(s)\neq t'$, then the language
      $Q(\overline{L})=
      \overline{\{i_1\cdots i_m w_{i_m}^R\cdots w_{i_1}^R@,~i_1\cdots i_m u_{i_m}^R\cdots u_{i_1}^R\# \mid m\ge 1\}}$, the language
    $P_{hi}(Q(\overline{L}))=
      \overline{\{i_1\cdots i_m w_{i_m}^R\cdots w_{i_1}^R \mid m\ge 1\}} \cup \overline{\{i_1\cdots i_m u_{i_m}^R\cdots u_{i_1}^R \mid m\ge 1\}}$, 
    and the assumption that the instance of PCP has no solution imply that $@$ and $\#$ are the only way to distinguish the strings $Q(s)$ and $t'$.
    If $Q(s)$ contains $@$, we may deduce that $s=i_1\cdots i_m \$ w_{i_m}^R\cdots w_{i_1}^R@$. Since $P_{hi}(t') = P_{hi}(Q(s)) = i_1\cdots i_m w_{i_m}^R\cdots w_{i_1}^R$, we have that $t'=i_1\cdots i_m w_{i_m}^R\cdots w_{i_1}^R \in Q(\overline{L})$. Such a $t'$ can be obtained either directly as $i_1\cdots i_m w_{i_m}^R\cdots w_{i_1}^R$ or as a strict prefix of the string $i_1\cdots i_m u_{i_m}^R\cdots u_{i_1}^R$. We now show that $s' = i_1\cdots i_m \$ w_{i_m}^R\cdots w_{i_1}^R$ satisfies MOC. Indeed, $Q(s')=i_1\cdots i_m w_{i_m}^R\cdots w_{i_1}^R = t'$ and $P(s) = i_1\cdots i_m \$ w_{i_m}^R\cdots w_{i_1}^R = P(s')$. Since the case where $Q(s)$ contains $\#$ is analogous, we conclude that if the instance of PCP has no solution, $\overline{L}$ is MOC, and hence OC by Lemma~\ref{MOCstrongerOC}.
  \end{IEEEproof}

\section*{Acknowledgment}
  The authors would like to thank the anonymous referees for their valuable comments and suggestions.

 \IEEEtriggeratref{19}

\bibliographystyle{IEEEtran}
\bibliography{biblio_journal}

\vfill

\end{document}